\documentclass[a4paper]{article}
\usepackage{a4wide}
\usepackage{url,microtype,textcomp}
\usepackage{authblk}
\usepackage{amsmath, amssymb}
\usepackage{wasysym}
\usepackage{lscape}
\usepackage{multirow}
\usepackage{booktabs}
\usepackage{xcolor}
\usepackage{tabularx}
\usepackage{colortbl}
\usepackage{siunitx}
\usepackage[numbers]{natbib}
\usepackage{geometry}
\usepackage{hyperref}
\newgeometry{margin=1in}
\usepackage[textsize=footnotesize]{todonotes}
\graphicspath{{Figures/}}
\definecolor{review1}{rgb}{0.0, 0.0, 0.0}
\definecolor{review2}{rgb}{0.0, 0.0, 0.0}

\newcolumntype{L}{>{\raggedright\let\newline\\\arraybackslash\hspace{0pt}}X}
\DeclareSIUnit[]{\no}{no}
\DeclareSIUnit[per-mode=fraction]{\npm}{\no\per\mm\squared}
\title{Computational Modeling of Cardiac Growth and Remodeling in Pressure Overloaded Hearts ---
  Linking Microstructure to Organ Phenotype}

\author[1]{Justyna A. Niestrawska}
\author[$1,\ast$]{Christoph~M.~Augustin}
\author[1,2]{Gernot Plank}
\affil[1]{\small Gottfried Schatz Research Center: Division of Biophysics,
          Medical University of Graz, 8010 Graz, Austria}
\affil[2]{\small BioTechMed-Graz, Austria}
\affil[$\ast$]{\small
    Correspondance:
    Christoph~M.~Augustin, Gottfried Schatz Research Center: Division of Biophysics,
  Medical University of Graz, Neue Stiftingtalstrasse 6/D04, 8010 Graz, Austria.
  christoph.augustin@medunigraz.at
}
\date{}

\begin{document}

\maketitle

% Abstract: recommended < 200 words (250 words max)
% Statement of significance: max 120 words
  \begin{abstract}
Cardiac growth and remodeling (G\&R) refers to structural changes in
myocardial tissue in response to chronic alterations in loading conditions.
One such condition is pressure overload
where elevated wall stresses stimulate the growth in cardiomyocyte thickness,
associated with a phenotype of concentric hypertrophy at the organ scale,
and promote fibrosis.
The initial hypertrophic response can be considered adaptive and beneficial
by favoring myocyte survival,
but over time if pressure overload conditions persist,
maladaptive mechanisms favoring cell death and fibrosis start to dominate,
ultimately mediating the transition towards an overt heart failure phenotype.
The underlying mechanisms linking biological factors at the myocyte level
to biomechanical factors at the systemic and organ level remain poorly understood.
Computational models of G\&R show high promise as a unique framework
for providing a quantitative link between myocardial stresses and strains at the organ scale
to biological regulatory processes at the cellular level which govern the hypertrophic response.
However, microstructurally motivated, rigorously validated computational models of G\&R
are still in their infancy.

This article provides an overview of the current state-of-the-art of computational models to study cardiac G\&R.
The microstructure and mechanosensing/mechanotransduction within cells of the myocardium is discussed and
quantitative data from previous experimental and clinical studies is summarized.
We conclude with a discussion of major challenges and possible directions of
future research that can advance the current state of cardiac G\&R
computational modeling.
\paragraph{Statement of Significance}
% overall significance of the problem
%Adaptation of the heart to pressure overloading by G\&R results in
%either adaptive or maladaptive patterns of hypertrophy.
%where over time maladaptive patterns dominate,
%initiating the transition towards heart failure (HF),
%a costly, progressive disease characterized by a significantly reduced quality of life and high mortality
%that is about to reach epidemic proportions in aging western societies.
%The pathobiology of adaptation processes is highly complex,
%and further complicated by the presence of various comorbidities
%(hypertension, diabetes, renal disease, atrial fibrillation, vasculopathy, metabolic syndrome).

% maybe we need only this part here, signficance of this review
The mechanistic links between organ-scale biomechanics and biological factors at the cellular size scale
remain poorly understood as these are largely elusive to investigations using experimental methodology alone.
Computational G\&R models show high promise to establish quantitative links
which allow more mechanistic insight into adaptation mechanisms
and may be used as a tool for stratifying the state and predict the progression of disease in the clinic.
This review provides a comprehensive overview of research in this domain
including a summary of experimental data.
Thus, this study may serve as a basis
for the further development of more advanced G\&R models
which are suitable for making clinical predictions on disease progression
or for testing hypotheses on pathogenic mechanisms using \emph{in-silico} models.

%parameterize computational models on the microstructural level and to develop a
%pathogenesis hypothesis suitable to be modeled \emph{in-silico}.

%This review provides a comprehensive overview of research
%on computational models of G\&R mechanism in the pressure overloaded heart,
%including experimental studies providing the data for model formulation and parameterization.
%Thus, this review serves as
%parameterize computational models on the microstructural level and to develop a
%pathogenesis hypothesis suitable to be modeled \emph{in-silico}.

\end{abstract}

%%% MAIN MANUSCRIPT %%%
\section{Introduction}

% General background and introduction to G&R
The heart is an electrically controlled mechanical pump
that propels blood through systemic and pulmonary \textcolor{review1}{circulations}.
It is able to adapt its output in various ways by changing rate or stroke volume
to match metabolic demands of peripheral tissues
over a broad range of conditions,
from sleep to exercise.
Its function depends on a highly complex and tightly orchestrated interaction
between a collection of cells (such as myocytes, fibroblasts, endothelial, and
vascular smooth muscle cells) and constituents of the extracellular matrix (ECM)
(mainly collagen, but also proteoglycans and some elastin).
The myocardium responds to changes in the environment,
such as alterations in the electrical activation sequences or in mechanical loading conditions,
by adaptation processes that alter myocardial structure and function.
These adaptation processes
which involve multiple molecular signaling pathways
are generally referred to as growth and remodeling (G\&R).
Prevalent processes implicated in most remodeling patterns comprise
altered myocyte dimensions
(such as changes in myocyte size and shape due to the addition of sarcomeres),
cell proliferation and apoptosis, and myocardial fibrosis.
Depending on the type of \textcolor{review1}{outcome}
processes can be classified either as adaptive when favoring myocyte survival
or as maladaptive when promoting apoptosis.
Overall, G\&R results in either adaptive or maladaptive patterns of hypertrophy
where adaptation during early states of a disease tends to be a compensatory beneficial mechanism
that normalizes heart function.
If pathological stimuli persist,
maladaptation increasingly dominates,
initiating a gradual progression towards more severe diseased states.

%which includes an elevated content of collagen, see~\cite{LeGrice2012b, Hill2008, Tanaka1986}.
%
%Growth is an initially compensatory mechanism to normalize the function of the
%heart in response to pathological stimuli such as pressure overload
%due to e.g.\ valve disease or hypertension.

% physiological background of pressure overload
Pressure overload of the left ventricle (LV) is a mechanical condition
that causes G\&R with a characteristic evolution of hypertrophic patterns \cite{Opie2006}.
The most prevalent pathologies causing pressure overload include
aortic stenosis, aortic coarctation or systemic hypertension.
These conditions are characterized by an elevated hemodynamic impedance
during LV systole.
Since cardiac output is governed by the metabolic demands of peripheral tissues
the needed hydrodynamic energy and, thus,
the biomechanical energy generated by the LV
for providing adequate cardiac output meeting these demands
increases with impedance of the systemic circulation.
\textcolor{review1}{
Thus the LV cavity must produce a higher pressure to compensate the pressure
losses due to the elevated impedance.
Higher pressure during systole translates then directly into an elevation in
afterload for the LV,
}
that is, the LV myocardium must produce higher systolic wall stresses,
but also indirectly to an increase in preload, i.e.\ increased wall stresses \textcolor{review1}{during diastole},
to maintain stroke volumes to produce the same normal cardiac output for a given heart rate.

% review currently prevailing hypothesis on adapation, stress correction and molecular adaptive vs maladaptive signaling
According to the systolic-stress-correction hypothesis \cite{grossman75:_stress_correction}
pressure overload stimulates myocytes to grow in width,
thereby increasing wall thickness and reducing wall stresses back to normal levels
as the increased myocardial wall volume facilitates the generation of the same chamber force at lower stress levels.
Thus LV hypertrophy (from the Greek word for increased growth) can be considered
a compensatory remodeling response,
leading to a typical concentric pattern of hypertrophy
characterized by an altered ratio of wall thickness to chamber volume.
Due to this adaptation the LV is able to maintain a normal cardiac output at a healthy ejection fraction (EF)
despite potentially impaired diastolic LV function
as a thicker walled potentially stiffer hypertrophied LV requires higher end-diastolic pressure
to achieve adequate filling.
At this mechanically compensated stage, pressure overload is usually asymptomatic.
However, over time if pressure overload is sustained,
end-diastolic pressure continues to rise, cardiac output and ejection fraction
start to decrease and the LV cavity dilates.
At the microstructural level,
progressive myocardial fibrosis and cell death takes place~\cite{Opie2006},
which leads to reduced ventricular compliance and progressive diastolic
dysfunction~\cite{Hein2003a, Krayenbuehl1989}.
Eventually, the initially compensatory, adaptive growth
associated with a concentric hypertrophic LV anatomy
transits into maladaptive growth associated with ventricular dilation,
initiating \textcolor{review1}{progress} towards overt heart failure.
It remains unclear whether molecular mechanisms account for this transition
-- from adaptive to maladaptive G\&R --
and, if so, whether interfering with signaling pathways can lead to new therapeutic approaches.
Studies on hypertrophic signaling pathways do not necessarily support the stress correction hypothesis
which is derived from macroscopic reasoning based on Laplace's law
\textcolor{review1}{\cite{esposito2002genetic,Heinzel2015,Lips2003}}.

While it appears obvious that any therapy must treat the root pathology causing the pressure overload condition,
this is, to a lesser extent, the case concerning the decision on optimal type and timing of an intervention.
For instance, aortic stenoses cause transvalvular pressure gradients leading to pressure overload in the LV.
Various clinical options exist to lower pressure gradients back to normal
such as surgical aortic valve repair or, for older or more frail patients, transcatheter aortic valve implant~\cite{Kheradvar2015,nishimura2017AHA}.
As those carry specific risks interventions are often postponed long after the asymptomatic phase
when first symptoms of heart failure start to manifest.
However, if interventions are carried out too late
the LV may have remodeled already to a significant degree
where the potential of beneficial reverse remodeling is substantially reduced.
\cite{McConkey2018,Everett2018,Otto2006,rahhab2019expanding}.

Also, there is a discrepancy between the onset of symptoms
and identified markers of long-term outcome after \textcolor{review1}{aortic valve replacement (AVR)}~\cite{Treibel2018}.
While reverse remodeling from a progressed state of disease to perfectly healthy conditions appears improbable,
a partial reversion of structural changes has been reported,
associated with marked reductions in LV mass and diastolic cavity size.
However, at a too progressed disease state the potential of myocardial plasticity appears exhausted~\cite{Yarbrough2012}.
Therefore, the search for biomarkers has been the target of current research
to find a better time point for surgical intervention,
i.e., the time point when adaptive remodeling turns maladaptive~\cite{Azevedo2010}.
Up until now there are no reliable criteria available to choose an optimal time point for an intervention
where the risk of progressing towards heart failure exceeds the risk of intervention
and at which remodeling is still largely reversible.

Similarly, the evaluation of the treatment's success poses challenges as well,
as the regression of fibrosis is progressing at slower time scales
than the regression of contractility \cite{Krayenbuehl1989}.
For instance, it has been reported \textcolor{review1}{by }\citet{Krayenbuehl1989}
that reverse remodeling remained incomplete
even \numrange{6}{7}~years after surgical aortic valve repair
as structural abnormalities still existed.
Thus, faster regression of contractility combined with slower regression of fibrosis
may lead to a LV with reduced compliance,
thus impairing diastolic filling and
-- due to insufficient preload --
the Frank-Starling mechanism,
and a limited contractile reserve.
Additionally, \citet{Krayenbuehl1989} hypothesized that this may point to a maladaption of the extracellular matrix (ECM) which manifested before aortic stenosis.

% start motivating this review
The underlying mechanisms driving these pathological processes occurring during pressure overload
are controversially debated \cite{Opie2006}.
While plausible hypotheses exist,
% we may add here some citations to the governing factors according to Opie et al, e.g. role of ratio MMP/TIMP; etc
neither the mechanisms by which cellular components at the molecular and microstructural level
respond to altered mechanical stimuli are fully understood \cite{Opie2006},
nor is the time-point known where adaptive changes turn maladaptive and irreversible.
%However, only with a thorough understanding future therapies may be improved.
As G\&R processes are governed by a number of pathways that interact in a complex manner,
computational modeling has been recognized as a promising tool
to quantitatively characterize the underlying mechanisms~\cite{Lee2016b}.
Such computational models usually combine a growth law based on a growth evolution model,
mostly adopting the approaches by~\citet{Skalak1996} and~\citet{Rodr1994a},
where growth was modeled by adding volume in prescribed directions.
\textcolor{review1}{While} conceptually simple, these models suffer from several disadvantages,
such as, e.g., the need of defining \emph{a priori} the direction in which growth will take place
or the limit at which growth will stop~\cite{vanOst2019a}.
Therefore, microstructurally motivated models in a framework such as constrained mixture theory
(see \textcolor{review1}{section }\ref{sec:CMT}) could provide deeper insight into the actual events leading to physiological
and pathological G\&R. To our knowledge, G\&R of the heart has not been modelled this way yet.

% outlook on content
This review paper aims to give an overview of the recent developments in the
dynamic field of G\&R with a specific focus on methodological aspects of modeling G\&R.
In the first part different concepts of modeling G\&R are reviewed.
Several excellent reviews have already covered the simulation of G\&R
\cite{Taber1995,Ambrosi2011,Ateshian2012,Menzel2012,Cyron2017,Wang2017,Witz2017a}.
However, to our knowledge a detailed summary of different modeling assumptions is lacking,
such as geometry, constitutive equations,
the driving factors of growth,
and the growth laws employed. \citet{Witz2017a} are the only ones to have systematically summarized,
compared and evaluated eight studies in the kinematic growth theory in a simple finite element setting.
To fill this gap we provide and discuss extensive summary tables
for the two most widely used G\&R frameworks,
kinematic growth theory, see~\autoref{sec:KGT},
and constrained mixture theory, see~\autoref{sec:CMT}.

The second part of the review provides a comprehensive overview
of all microstructural components of the heart
which might be important to be understood and considered from a engineering point of view.
Only then, a detailed description of the changes in microstructure in pressure overloaded hearts is given
which aims to provide a common ground based on which the pathogenesis might be modeled.
An extensive table summarizes quantitative findings on changes in microstructure
from studies of the last \num{30} years which utilized biopsies from surgeries
at different time points before, during, or after AVR\@.
As detailed quantitative data \textcolor{review1}{are} essential for sound model development
and validation, we believe that this review might serve as a motivation and a
great help to extend the already growing field of phenomenological G\&R models
to the use of constrained mixture theory.
This might enhance the understanding of multiscale mechanisms
responsible for the progression towards heart failure or for full recovery post-treatment
which may translate eventually into clinically applicable modeling tools
for stratifying the disease state and predicting its progression and the potential for reverse remodeling.
As there has been a considerable debate about
which growth stimuli \textcolor{review1}{are} being sensed by cells,
a final subchapter is devoted to mechanosensing and mechanotransduction in cells.

%%%%%%%%%%%%%%%%%%%%%%%%%%%%%%%%%%%%%%%%%%%%%%%%%%%%%%%%%%%%%%%%%%%%%%%%%%%%%%%
\section{Computational Treatment of Growth and Remodeling}
Several models to study G\&R computationally have been developed in the past, such as
\textcolor{review1}{a recently proposed framework of "relaxed growth" by \cite{Genet2015a}, simple lumped parameter models to study growth stimuli \cite{Rond2019a}} or
the theory of porous media \cite{Hopk2018a}.
However, we \textcolor{review2}{first} concentrate on the two most widely used approaches,
namely kinematic growth theory and constrained mixture models \textcolor{review2}{and give a summary of structural adaptation theory and fully structural theory, as the latter proposes a solution to the problem of the assumption of bijectivity, see chapter \ref{sec:Struc}.}
We then give a short summary on hybrid approaches and conclude with a summary
of strain-energy functions used in the literature.
%
%%%%%%%%%%%%%%%%%%%%%%%%%%%%%%%%%%%%%%%%%%%%%%%%%%%%%%%%%%%%%%%%%%%%%%%%%%%%%%%
\subsection{Kinematic Growth Theory}\label{sec:KGT}
\begin{figure}[ht]
  \centering
  \includegraphics{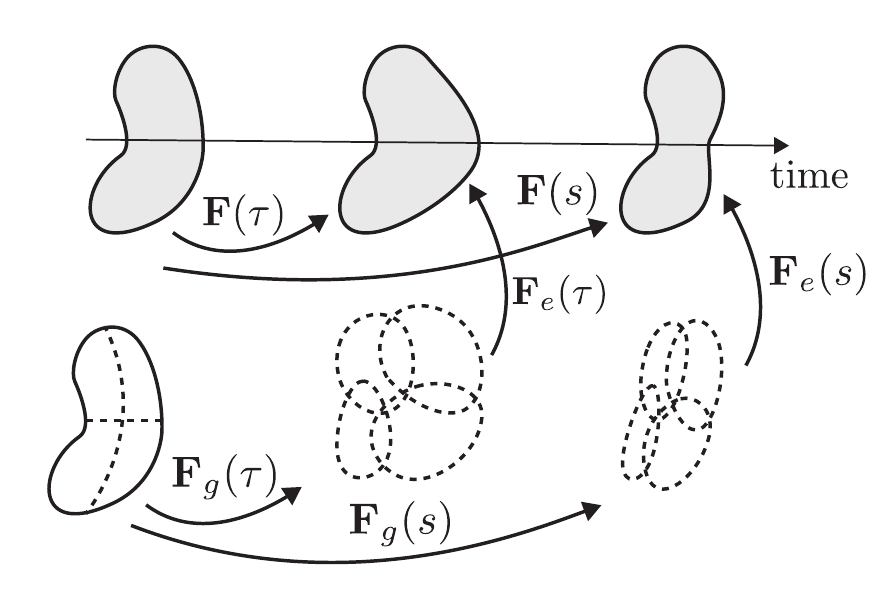}
  \caption{In kinematic growth theory, a body is deformed due to growth and
    external loads in two time points $\tau$ and $s$. The total deformation
    gradient is decomposed into an inelastic growth part $\mathbf{F}_{\rm g}$ and
    an elastic part $\mathbf{F}_{\rm e}$, leading to geometric compatibility
and mechanical equilibrium. \textcolor{review2}{Adapted with permission from}~\citet{Cyron2017}.}%
\label{fig:Fig1}
\end{figure}

Based on the concept of plasticity \cite{Lee1969} the construct of volumetric
growth in a continuum formulation was first proposed by \citet{Rodr1994a}.
Growth is modeled there as a change in shape and size of an unloaded body by
means of the inelastic growth deformation gradient $\mathbf{F}_{\rm g}$.
Stress-free changes of volume elements are described due to added or lost mass,
leading by itself to a not compatible, intermediate configuration,
see~\autoref{fig:Fig1}.
In a second step, geometrical compatibility is achieved by the elastic deformation
gradient $\mathbf{F}_{\rm e}$,
which assembles the volume elements into an unloaded body,
leading to the development of residual stresses.
Overall, the total deformation gradient results in
\begin{equation}
\mathbf{F} = \mathbf{F}_{\rm e}\mathbf{F}_{\rm g}.
\label{eq:KinDefo}
\end{equation}
This approach leads to two separate constitutive relations:
an evolution equation for the growth tensor and a
strain-energy density function, dependent only on the elastic deformation as
$\Psi(s)=\Psi(\mathbf{F}_{\rm e})$ at any time $s$.
While this approach is conceptually simple, the definition of the growth law remains a major challenge.

The concept of kinematic growth has been utilized for the computation of
patient-specific geometries of the aorta in several studies
\cite{Rach1998a,Tabe1998a,Alas2007b,Kuhl2007,Rodr2007b, Saez2014a}.
The first to use this concept to model G\&R in the heart were
\citet{Kroon2009a}, using a simple truncated ellipsoid geometry and a
transversely isotropic material model.
They investigated the impact of two different reference configurations:
a single one which remains fixed throughout the entire growth process,
and one which is updated after each growth increment. They concluded that the
choice of the reference configuration is of great importance as it has
a significant impact on the growth stimulus.
The biomechanical driving factors of growth \textcolor{review1}{are still} unknown.
However, the current prevailing hypothesis states that stress is the driving factor for
concentric cardiac hypertrophy due to pressure overload\textcolor{review1}{,} and strain drives
eccentric cardiac hypertrophy due to volume overload.
Both stimuli were incorporated in the works by~\citet{Goek2010a,Gokt2010b}
that described both eccentric and concentric growth.
They hypothesized that eccentric growth is due to serial sarcomere deposition,
whereas parallel sarcomere deposition leads to concentric growth.
ECM remodeling was neglected in these studies, and a phenomenological critical
myofiber stretch was used,
from which an excess would trigger eccentric remodeling.
Only passive material parameters were used and simulations started from
an unloaded state. Several studies followed the approach, such as,
e.g., \citet{Lee2015} where reverse growth occurred when myofiber stretch fell below
a prescribed homeostatic value.
An isotropic growth tensor was used to study the influence
of the chosen growth multiplier. The growth multiplier was modeled in such a way
that reverse growth was possible when the stretch fell below the threshold of
a critical fiber stretch which initiated growth in the first place.
The model was integrated into an electromechanical model of the heart
\cite{Sundnes2014} \textcolor{review1}{to model active states of the myocardium as well} in a subsequent study \cite{Lee2016}.
In 2011, \citet{Rausch2011a} investigated the mechanisms by which the heart
remodels due to pressure-induced sarcomerogenesis.
The contribution of active stresses \textcolor{review1}{was} introduced by modifying equations due to
~\citet{Gucc1993b}. \citet{Klepach2012a} examined the influence of myocardial
infarction on the shape of the heart, inducing heterogeneity by different
(passive) material parameters for infarct and healthy regions.
\citet{Kerckhoffs2012} gave a detailed explanation on their choice of
biomechanical stimuli, resulting in a more complicated growth law,
where the relation between the orthotropic components of the growth tensor and
the stimuli was incremental and described by sigmoidal functions.
\textcolor{review1}{It} was assumed that all adaptation mechanisms were present under all conditions,
i.e., while sarcomeres are mostly added in parallel in pressure overload
and in series in volume overload,
both adaptation mechanisms should be present independent of the driving conditions,
but to different extents.
The homeostatic value for fiber and cross-fiber strain was
set to $0$.
\citet{Genet2016a} applied kinematic growth to a four-chamber human heart model, examining the question whether computational modeling can
provide patient-specific information about the progression of heart failure.
Recently, \citet{Witzenburg2018} coupled the growth law proposed by \citet{Kerckhoffs2012}
to a time-varying elastance compartmental model, which was connected to a
lumped model of the circulation.
In their extensive study,
multiple \textcolor{review1}{experimental} results were reproduced by simulation of volumetric
and pressure overload\textcolor{review1}{,}
and the ability of the model to predict growth in three studies of pressure and
volume overload as well as post-infarction remodeling was examined.
Despite using a phenomenological growth law, they were able to predict results
from independent studies of all loading conditions and claimed that their model
was able to automatically customize model parameters and to calculate
3 months of G\&R in 6 minutes simulation time.
In 2018, \citet{DelBianco2018} investigated how hypertrophy affects the heart's
electrophysiology.
They assumed that growth depends on multiple stimuli throughout the cardiac cycle
as opposed to just one stimulus from one specific point in the cycle. Hence
they modelled growth to occur at the end of each cycle.
The most recent study was performed by \citet{Peir2019a}, utilizing machine learning
and information from multiple scales obtained from a porcine model of volume overload.

An overview on studies utilizing the kinematic growth approach to model G\&R
in the heart is given in~\autoref{tab:KinGr},
where the used growth stimuli, growth laws, and geometries are summarized.

Further work, which is not included in~\autoref{tab:KinGr},
has been done by Taber~et~al.~\cite{Lin1995,Tabe2002a} who used the framework
of kinematic growth theory to describe the growth of a developing heart.
Similar to \citet{Kroon2009a} and \citet{Kerckhoffs2012a} they utilized the
excess of the stress tensor from its target value as the growth stimulus.
As the incompatibility of the growth tensor induces naturally residual stresses
and strains, the framework has been also used to model residual stresses
by \citet{Skalak1996} in general and more recently for the heart by \citet{Genet2015a}.

As each of the discussed studies not only utilizes different growth laws,
but also different geometries, (passive) constitutive equations and loading
conditions, \citet{Witz2017a} systematically compared eight kinematic
growth laws. They conducted a literature review \textcolor{review1}{of} different surgical
procedures and summarized a range of change of stretches induced by them.
These data were then used as an input for comparison of the different growth
laws and compared to the amount actual growth reported from experimental studies.
They concluded that only laws using multiple inputs which were weakly correlated
with one another in experimental studies were able to capture features of
both pressure- and volume-overload. However, none of the tested models were
able to return to homeostasis in prescribed-force simulations, but some were in
prescribed-stretch simulations. \citet{Witz2017a} summarized a range
of data they gathered for their study and found that in pressure overload,
end-systolic fiber stretch was increased and volume overload \textcolor{review1}{it was decreased}, whereas
end-diastolic stretch (often used as an input in kinematic growth laws) increased
in both overload scenarios. Hence they concluded that it
should be possible to capture pressure and volume overload with a single input
connected to end-systolic fiber stretch.

Kinematic growth theory is a conceptually simple and computationally
convenient method, which is, however, limited mechanobiologically~\cite{Cyron2017}.
Initially, the theory was designed to model growth of stress-free configurations.
Yet, living tissues such as the heart are pre-stressed and consist of
multiple constituents whose removal, deposition, and adaptation governs the
stress state inside the body. Especially during pathogenesis the disturbed
deposition and removal of constituents such as collagen has a significant
impact on the mechanics~\cite{King2013,Nguyen2014}, which cannot be captured
by a single phenomenological evolution law of $\mathbf{F}_{\rm g}$.
Additionally, the spatial density $\rho$ is often assumed to remain constant
during both elastic deformation and G\&R. \textcolor{review1}{In summary}, kinematic
growth theory is able to capture the consequences of growth, but not the
processes which lead to it.
%

%%%%%%%%%%%%%%%%%%%%%%%%%%%%%%%%%%%%%%%%%%%%%%%%%%%%%%%%%%%%%%%%%%%%%%%%%%%%%%%
\subsection{Constrained Mixture Theory}\label{sec:CMT}
A fundamentally different approach are constrained mixture models, first
proposed in the context of biomechanics by \citet{Hump2002a},
which are based on the theory by \citet{True1965a} of interdiffusing,
non-reactive materials. Here, G\&R is attributed to the
deposition and degradation of individual constituents within a tissue.
The main hypothesis is that cells can synthesize new components of a
constituent $i$ which are deposited in preferred directions at time
$\tau \in [0,s]$ with a certain pre-stretch $\mathbf{F}^{i}_\mathrm{pre}(\tau)$.
This stretch can be interpreted as the homeostatic stretch imposed on a
fiber that is incorporated within the ECM.
At the same time mechanisms such as cell apoptosis or matrix degeneration can
lead to the degradation of components.
Let $\mathbf{F}(s)$ and $\mathbf{F}(\tau)$ denote the
deformation gradients that account for motion of the whole continuum from the
reference configuration at time $s=0$ to configurations at times $s$ and $\tau$,
respectively. Then, the product $\mathbf{F}(s)\mathbf{F}^{-1}(\tau)$ represents
the deformation of the whole continuum from time $\tau$ to the current time $s$.
Then the individual deformation gradient for a constituent $i$ that
describes the deformation from its natural, stress-free state at time of
production $\tau$, emphasized by the superscript $i(\tau)$,
to the current time $s$ results in
\begin{equation}
  \mathbf{F}^{i(\tau)}_{\mathrm{e}}(s)
  = \mathbf{F}(s)\mathbf{F}^{-1}(\tau)\mathbf{F}^{i}_\mathrm{pre}(\tau).
\label{eq:MixF}
\end{equation}
Further, the individual right Cauchy--Green tensor of each cohort of each
constituent $i$ is defined as
\begin{equation}
  \mathbf{C}^{i(\tau)}_{\mathrm{e}}(s) =
    \left(\mathbf{F}^{i(\tau)}_{\mathrm{e}}(s)\right)^\top
    \mathbf{F}^{i(\tau)}_{\mathrm{e}}(s).
\end{equation}
Constrained mixture models assume that all constituents, regardless of
their time of deposition or their individual pre-stretch, deform together when
subjected to an external load.

The existence of multiple constituents, deposited at different times and under
different pre-stretches, \textcolor{review1}{gives} rise to residual stresses, without incompatible
growth strains as assumed in kinematic growth theory~\cite{Ateshian2012}.

The concept of constrained mixture theory basically leads to three types of
constitutive relations which have to be defined:
\begin{enumerate}
  \item the stored energy function $\hat{\Psi}^i$ for a constituent $i$,
    depending on the deformation experienced at the current time $s$,
    relative to its natural configuration at the G\&R time of
    its production $\tau$;
  \item the rate of production of mass per unit volume $m^i(\tau)$; and
  \item the associated survival function $q^i(s,\tau) \in [0,1]$,
    describing the half-life of each constituent, i.e., the percentage of
    constituent $i$, deposited at time $\tau$, and surviving to
    the current time $s$ \cite{Ateshian2012}.
\end{enumerate}
The total stored energy density function $\Psi^{\rm{tot}} = \sum_{i=1}^{n}\Psi^{i}$,
where $n$ is the number of constituents, is usually calculated using a mass
average rule, such as, e.g., in~\cite{Vale2013a}:
\begin{equation}
  \Psi^{i}(s)=\frac{\rho^{i}(0)}{\rho(s)}q^{i}(s,0)
    \hat{\Psi}^{i}(\mathbf{C}^{i(0)}_{\mathrm{e}}(s))
     + \int_{0}^{s}\frac{m^{i}(\tau)}{\rho(s)}q^{i}(s,\tau)
       \hat{\Psi}^{i}(\mathbf{C}^{i(\tau)}_{\mathrm{e}}(s))d\tau,
\end{equation}
  where $\rho^{i}(0)$ is the mass density of constituent $i$ at time $s=0$ and
   $\rho(s)$ equals the total mass density.
The production rate $m^{i}(\tau)$ is the difference between production and
degradation of constituent $i$ and often assumed to be stress-dependent.
Additionally, survival functions are often chosen to take an exponential form.

Applications of constrained mixture models
to cardiac modeling problems, to the best of our knowledge,
have not been reported yet. The bulk of the modeling work in the literature focused on
vascular applications. The state-of-the-art in vascular constrained mixture modeling
has been summarized in reviews by \citet{Ateshian2012}, who showed
examples of the diverse applications of mixture theory in biological growth
and remodeling and discussing open problems. At the same time,~\citet{Vale2012a}
reviewed ten years of constrained mixture models, summarized the core hypotheses
in these studies and briefly recapitulated the core mathematical concepts used,
mostly focusing on arterial G\&R. The majority of publications in this field
utilized 2-D membrane models in which volume growth was incorporated through
a change in membrane thickness,
e.g.,~\cite{Glea2004a,Kroo2007a,Baek2005b,Vale2009b,Figueroa2009},
often assuming a mean intramural stress. The formulation was extended to
3-D by, e.g., \citet{Karsaj2010}, who used a semi-analytical solution of the
extension and distension of a straight cylindrical segment of an artery.

Constrained mixture models were used within the finite element framework by,
e.g., \citet{Machyshyn2010}, who simulated isotropic growth in an externally
unloaded state and assumed stretch as a growth stimulus. The growth-induced
stress was assumed to disappear due to tissue remodeling. Further studies
assuming isotopic growth kinematics were conducted in the following years
\cite{Vale2013a,Schmid2012,Eriksson2014}, however, unrealistic model
predictions were reported by \citet{Vale2013a}, leading to models employing
anisotropic model kinematics, e.g.,~\citet{Wan2010} for orthotropy.
Anisotropic model kinematics can be uniquely defined by eigenvectors and
eigenvalues of the deformation due to growth, as earlier studies showed for
transversely isotropic growth kinematics, e.g.,~\citet{Baek2005a}.

\autoref{tab:ConsGr} summarizes constitutive equations and assumptions used
by exemplary studies
% The "after" here is meant in a temporal sense or rather means "according to"? Don't use the lengthy references over and over again, you have done this above, vanity is not a scientific category, you  don't need to name Humphrey and Holzapfel over and over again to please their ego's.
% Justyna: Temporal sense, what should I use instead to make this clear? Humphrey didn't seem vain at all talking to him personally, just as a side note ;-).
published after the reviews of \cite{Ateshian2012}
and \cite{Vale2012a}, utilizing the classical constrained mixture theory or a
hybrid approach, discussed in the following chapter. In 2013, \citet{Vale2013a}
simulated the growth of an aneurysm within a 3D framework by an exponential
decay of the original shear modulus of elastin. They analyzed the numerical
stability of their implementation and concluded that for their case of an
axisymmetric problem, at least 10 half-lives of collagen had to be tracked in
order to ensure numerical stability. \citet{Wu2015} modeled aneurysm
development in a membrane model of a patient-specific infrarenal aorta,
coupling G\&R with blood flow simulations.
In the same year, \citet{Virag2015} extended the model of aneurysm development
by \citet{Karsaj2010} by incorporating the evolution of multiple layers of an
intraluminal thrombus. More recently, \citet{Famaey2018} studied the influence
of the Ross procedure on arterial wall mechanics, where the pulmonary artery
is exposed to a seven fold increase in blood pressure.
They studied four virtual tests using one single element with boundary
conditions assigned using Laplace's law. Finally, in 2019, \citet{Horvat2019}
studied the influence of the volumetric-isochoric split of the strain-energy
function on G\&R in a 3D finite element implementation of a
three-layered aorta.

\citet{Latorre2018} proposed a mechanobiologically equilibrated, steady-state
formulation, valid for states in which the tissue has completed G\&R
due to long term external stimuli.
In the case of a thin-walled artery the steady-state formulation yielded the
exact same solution as a full constrained mixture model.

Other applications of the constrained mixture theory are, e.g., the definition
of constitutive laws based on constituent-specific pre-stretches, leading to
residual stresses as introduced by \citet{Mousavi2017}.
This can be used to understand the influence of sample preparation before
uniaxial or biaxial tensile testing on the microstructure and hence on the
mechanical behavior.
%
%%%%%%%%%%%%%%%%%%%%%%%%%%%%%%%%%%%%%%%%%%%%%%%%%%%%%%%%%%%%%%%%%%%%%%%%%%%%%%%
\subsection{\textcolor{review2}{Structural Adaptation Theory}}\label{sec:StrucAdap}
\textcolor{review2}{As mentioned in chapter \ref{sec:KGT} kinametic growth
	theory is able to model the consequences of growth, but not the processes
	during growth. A group of models to include structural reorientation was
	first described by \citet{Arts1979}, where the left ventricle was modeled by a set
	of concentric cylinders. The model yielded a transmural variation of myofiber
	orientation similar to experimental observations. The orientation as well as
	the wall thickness were adapted by optimization of the local deviation of the fiber shortening during systole and the sarcomere length at the beginning of the ejection from a set
	value. It was assumed that remodeling occurs due to continuous creation and breaking
	down of connections between fibers and their environments. The model predicted
    a homogeneous fiber stress and strain distribution during the ejection phase.
    This finding was then translated into the 'one fiber model'  which approximated
	fiber stress and strain by single values throughout the whole wall \cite{Arts1991}.
    The model approximated LV cavity volume and LV pressure by relating them to myofiber
    stress and stretch through a simple law. The model by \citet{Arts1979} was extended with rules for
    adaptation to account for cardiac growth by \citet{Arts1994}. }

\textcolor{review2}{In 2005, \citet{Arts2005} proposed an adaptation model named "CircAdapt". Here, two one fiber models for left and right ventricular function together with a lumped
parameter model of the circulation were utilized to simulate patient hemodynamics.
Cavity volume was increased in case sarcomere shortening during the ejection exceeded a
predefined amount of shortening and wall volume increased when sarcomere length exceeded
a predefined length at end diastole. The model was able to predict adaptive changes of
the heart three months before until three months after birth. }

\textcolor{review2}{\citet{Lumens2009} extended this framework to model the interaction between left and
right ventricle more realistically, allowing a separate description of mechanics and adaptation
of free walls and septum of the ventricles.}

\textcolor{review2}{Kroon et al. \cite{Kroon2009} modeled the adaptive reorientation of cardiac
myofibers within the framework proposed by Arts and colleagues, prescribing evolution equations for the structural
tensors characterizing anisotropic constitutive parameters. The remodeling law stated
that myofibers adapt their direction in order to minimize the shear strain between the fiber and
cross-fiber directions.}

\textcolor{review2}{In a subsequent study, Pluijmert et al. \cite{Pluijmert2012} examined the
influence of initial and boundary conditions on the shear induced reorientation of fibers.
They concluded that physiological boundary conditions are crucial, however,
as the available experimental data shows a large disparity, adequate
boundary conditions are yet to be defined.}

\textcolor{review2}{Consequently, \citet{Pluijmert2017} utilized the
previously established model to examine the influence of myofiber orientation in a biventricular
finite element model on the simulated cardiac function. They used their model to alter the fiber
direction by letting the myofibers remodel and with that to define the myofiber orientations
 in finite element models.}

\textcolor{review2}{For an excellent review and more detailed discussion on this class of
	models the reader is referred to \cite{Bove2012b}. Structural adaptation models are a useful
tool to study vascular adaptation with computationally cheap models. However, if the spatial distribution of myocardial growth is of interest, one has to utilize the above mentioned
continuum mechanical formulations and couple models such as the one proposed by  \citet{Pluijmert2017} to growth laws.}
%
%%%%%%%%%%%%%%%%%%%%%%%%%%%%%%%%%%%%%%%%%%%%%%%%%%%%%%%%%%%%%%%%%%%%%%%%%%%%%%%
\subsection{\textcolor{review2}{Fully Structural Approach}}\label{sec:Struc}

\textcolor{review2}{One of the major drawbacks of the above mentioned theories which utilize
	the multiplicative decomposition of the deformation gradient into an elastic and a growth
	part is the assumption that the motion is bijective. However, \citet{Cowin2010} pointed out that finite volumetric growth including mass change is not bijective, i.e., a material point can disappear by degradation or grow and become multiple points. Thus, there is no one-to-one mapping in tissue growth which includes mass change. In fact, due to this non-bijectivity, the definition of kinematics of finite volumetric growth with mass change is yet an unresolved problem in continuum mechanics.}

\textcolor{review2}{The only approach aiming to overcome this issue was recently
proposed by \citet{Lanir2015a,Lanir2017a}. Here, in the earlier study a one
dimensional G\&R theory was proposed and extended to three dimensions
afterwards.}

\textcolor{review2}{The multi-scale approach is linking turnover events occurring at
	the constituents' level with the evolving tissue structure. The
structural approach models growth as the adaptation of fiber bundles by
change of the number of fibers in each bundle. As the bundles themselves
cannot disappear and no new bundles are formed the deformation of a bundle is
affine with the global tissue --- with this assumption bijectivity of the
model is guaranteed. \cite{Lanir2017a}}

%\textcolor{review2}{The main difference to constrained mixture theory is that there, the structure
%is not considered and hence the tissue's remodeled configuration is impossible
%to be determined in 3D. \cite{Lanir2017a}}

\textcolor{review2}{While this approach seems promising, it is yet restricted to
purely collagenous tissues such as tendons and ligaments and has yet to be developed
further to be able to be used for complicated multi component tissues such as the heart.}

%
%%%%%%%%%%%%%%%%%%%%%%%%%%%%%%%%%%%%%%%%%%%%%%%%%%%%%%%%%%%%%%%%%%%%%%%%%%%%%%%
\subsection{Hybrid Approaches}
%
%%%%%%%%%%%%%%%%%%%%%%%%%%%%%%%%%%%%%%%%%%%%%%%%%%%%%%%%%%%%%%%%%%%%%%%%%%%%%%%
\subsubsection{Volumetric Constrained Mixture Models, Evolving Recruitment Stretch}

\citet{Alfo2008a} were the first to model thick-walled aortic
geometries within the framework of constrained mixture theory, combining it
with kinematic growth theory to test the influence of several factors,
such as elastin distribution, axial pre-stretch, and the effect of wall
constituents on the resulting opening angles. \citet{Wan2010} also utilized an
analytical model to  model a thick-walled aorta, calculating local homeostatic
pre-stretches with pre-determined distribution functions.
A series of studies was published by Watton and co-workers
\cite{Watt2004a,Watt2009c,Watt2009b,Watt2011a,Eriksson2014,Selimovic2013a},
utilizing the constrained mixture theory to model aneurysm enlargement.
Instead of tracking different reference configurations of mass increments
deposited at different times, they focused on the change in the stress-free
configuration by tracking the evolution of a recruitment stretch in an
intermediate configuration. In 2017, \citet{Grytsan2017} extended this framework
to study different cases of volume growth assumptions. They found that
assuming isotropic or in-plane growth yielded unphysiological results
and should be used with caution in the case of aneurysm growth. Recently,
Lin~et~al.~\cite{Lin2018, Lin2019} simulated the dilatation of a thick-walled
aneurysm. They assumed that the dilatation is due to a local degradation of
elastin, accompanied by a stretch-mediated degradation of collagen and an
increased production of new collagen, which is deposited dependent on a
recruitment variable defining an intermediate configuration where collagen
fibers start to bear load. Despite yielding promising results, they
encountered a prominent thinning of the vessel wall, which is not physiological.

Finally, it is worth noting that~\citet{Schmid2012} implemented a 3D finite
element model of a hybrid approach \textcolor{review1}{based in part} on the constrained mixture theory, but instead of
being integral based, their approach was rate based.

%%%%%%%%%%%%%%%%%%%%%%%%%%%%%%%%%%%%%%%%%%%%%%%%%%%%%%%%%%%%%%%%%%%%%%%%%%%%%%%
\subsubsection{Homogenized Constrained Mixture Theory}
A new model has been introduced by Cyron et al.\ to avoid keeping
track of all past tissue configurations while still being able to account
for micro-structural influences. The main difference between constrained
mixture and the proposed theory is how mass production and volumetric growth are
computed. An inelastic G\&R deformation $\mathbf{F}^i_\mathrm{gr}$ is treated as the
product of exchange-related remodeling $\mathbf{F}^i_{{\rm r}}$ and growth
related deformation $\mathbf{F}^i_{{\rm g}}$ for each constituent $i$ resulting
in the total deformation
$\mathbf{F}=\mathbf{F}^i_\mathrm{e}\mathbf{F}^i_\mathrm{gr}
 = \mathbf{F}^i_\mathrm{e}\mathbf{F}^i_\mathrm{r}\mathbf{F}^i_\mathrm{g}$,
~\cite{Cyron2016}.
Here, the specific growth law has to be assigned similar to kinematic growth
theory. A comparison to established constrained mixture
theory based simulations showed marginal differences in the outcome.

Recently, \citet{Braeu2018} used the concept of tensional homeostasis
(i.e.~the assumption that cells remodel in a way to maintain a preferred state
of stress) as a driving factor for anisotropic growth. This enabled them to
refrain from using a phenomenological growth tensor. Their study was based on
two hypotheses stating
1.~that tensional homeostasis is the only driving factor for growth and
2.~that the mass density is constant during growth.

%

%\subsection{Structural Adaptation}

%\textcolor{review1}{write later or include at all?}

%2005 Arts et al Circ adapt model: Combination of active sarcomere mechanics model,
%analytical expressions of fiber stretch and stress relation to cavity pressures and volumes,
%circuit model of corculation full cardiac cycle simulation before and during pressure overload.
%max to min fiber stretch radial growth law

%Kroon et al. \cite{Kroon2009} modeled the adaptive reorientation of cardiac
%myofibers within this framework, prescribing evolution equations for the structural
%tensors characterizing anisotropic constitutive parameters.
%In a subsequent study, Pluijmert et al. \cite{Pluijmert2012} examined the
%influence of initial and boundary conditions on the reorientation of fibers.

%
%%%%%%%%%%%%%%%%%%%%%%%%%%%%%%%%%%%%%%%%%%%%%%%%%%%%%%%%%%%%%%%%%%%%%%%%%%%%%%%
\subsection{Strain-Energy Functions}
\citet{Dokos2002a} performed simple shear experiments on dog hearts
that revealed, for the first time, hyperelastic anisotropic behavior of
the myocardium. This behaviour has been ever since represented based on Fung-type
exponential functions. Different approaches have been chosen, such as modeling
the heart as a transversely anisotropic~\cite{Gucc1991a} or
orthotropic material~\cite{Cost1999b}.

Similar shear experiments have been repeated on
human myocardium by Sommer~et~al.~which revealed pronounced viscous
properties in addition to the previously reported orthotropic behavior~\cite{Somm2015b}.
However, there are no data available representing the whole heart and hence
hyperelastic models still remain standard.

In 1994~\citet{Nevo1994a} modeled the passive stress response utilizing
exponential functions coupled to a undulation density distribution which
summed the contribution of individual collagen fibers.
They stated however that the quantification of the collagen volume fraction
is difficult in the heart.

Later,~\citet{Holz2009c} proposed a microstructurally motivated orthotropic
model based on Fung-type exponential functions
which was later augmented by incorporating fiber dispersion \cite{Erik2013b}. Following more
recent experimental findings on the viscoelastic behavior, orthotropic
viscoelastic material models have been proposed in~\cite{Gult2016a,Cans2015a}.
\citet{Wang2014a} have modified the model by~\citet{Holz2009c} to model tissue
remodeling taking place in hypertensive heart failure in rats utilizing
image-derived structural parameters representing the growth of perimysial and
endomysial collagen.

Using these equations to describe cardiac remodeling remains a challenge
due to their phenomenological nature and parameterization, especially
\emph{in-vivo}. Additionally, the microstructure of the heart is not as easily
characterized as for other soft tissues such as blood vessels or skin and there is
still a profound lack of available data~\cite{Wang2015}.

In 2017,~\citet{Avaz2017} used microstructural parameters quantified from
histological studies to characterize the passive response of the right
ventricle with a constitutive equation on the fiber level, separating
mechanical contributions of myofibers and collagen and modeling the connection
between these. Aspects of tissue adaptation have been modeled by,
e.g.,~\cite{Drie2005a,Hari2007a}, prescribing evolving constitutive equations
on the fiber level. Others utilized the adaptation of material parameters to
model changes in the microstructure~\cite{Wang2016a} or added new
parameters to introduce maximum energy accumulated before
failure~\cite{Volo2008b}. Although these models provided some insight into
the changes taking place during the pathogenesis of various diseases by
changing the constitutive behavior, none of them is able to capture volume
changes due to changes in mass. Neither did any of the studies account for changes
such as myocyte hypertrophy. A summary table on constitutive equations for the
passive myocardium is given in the excellent review by~\citet{Wang2017}.
%
%%%%%%%%%%%%%%%%%%%%%%%%%%%%%%%%%%%%%%%%%%%%%%%%%%%%%%%%%%%%%%%%%%%%%%%%%%%%%%%
\section{Biological Processes of Growth and Remodeling During Pressure Overload}
Over the past \num{30} years extensive research has been dedicated to
studying micro-structural remodeling under pressure overload conditions.
Experimental studies relied mostly on animal models
where pressure overload was induced by aortic banding using aortic cuffs (e.g. \textcolor{review1}{\cite{Omen1996a,Nedi2000a,Naga2000a,Schu2001a,Moor2006a,Yarb2012a}}\cite{Omen1996a,Nedi2000a,Naga2000a,Schu2001a,Moor2006a,Yarb2012a} for large animals)
or, genetically modified mouse or rat models were employed
that develop diastolic heart failure and subsequently, if left untreated, systolic heart failure\textcolor{review1}{, which resembles the disease progression seen under pressure overload.}
(e.g. \cite{LeGrice2012b, Wils2017a}).

%\todo{citation aortic compliance and wall stresses needed}
%\begin{itemize}
%	\item rodent myocardium short action potential lacking a plateau phase
%	\item resting heart rate 5 times that of human
%	\item force-frequency relation is inverse (in humans positive)

%	\item in larger animals, like dogs the beta myosin heavy chain isoform is predominant
%	\item excitation-contraction coupling similar to human in contrast to rodents

%	\item review on animal models of human heart failure \cite{Hase1998a}
%\end{itemize}

The procedure of aortic banding, however, is afflicted with uncertainties.
For instance, it has been shown that even minor differences in cuff placement
may result in significantly different end-systolic pressure
(e.g. \SI{18.1}{\mmHg} in~\cite{Crozatier1984} and
\SI{38}{\mmHg} in~\cite{Sasayama1977} in the same animal model).

As it is not possible to study the onset of the disease in patients,
animal studies remain a valuable source of information,
but caution is warranted when extrapolating to human conditions, not only because
of inter-species differences but also because of differences in experimental procedures.

Due to the limited space in this review this section therefore focusses on quantitative data gained in
clinical studies over the past three decades. These data can can serve as quantitative input to the
formulation of constitutive G\&R laws.
% on the micro-structural model.
Further, an overview on the microstructural \textcolor{review1}{composition} of the LV myocardium in health and disease is given.
The sections concludes with a summary of the pathogenesis and an overview over how
cells such as fibroblasts or myocytes can sense and react to mechanical stimuli in their
environment.

%%%%%%%%%%%%%%%%%%%%%%%%%%%%%%%%%%%%%%%%%%%%%%%%%%%%%%%%%%%%%%%%%%%%%%%%%%%%%%%
\subsection{Heart Constituents under Physiological Conditions}

%%%%%%%%%%%%%%%%%%%%%%%%%%%%%%%%%%%%%%%%%%%%%%%%%%%%%%%%%%%%%%%%%%%%%%%%%%%%%%%
\subsubsection{Myocytes}

Although only about one third of all cells making up the myocardium are cardiomyocytes,
they account for \SI{75}{\%} of its volume \textcolor{review1}{\cite{Pagel2018}}.
Cardiomyocytes are striated muscle cells with a diameter of around \SI{25}{\um}
and a length of $\approx$\SI{100}{\um} \textcolor{review1}{\cite{Pagel2018}}.
While myocytes are single nucleated,
from a functional point of view they can be viewed as a ``syncytium''
as they form tight connection to neighboring cells
via specialized cell membranes referred to as intercalated disks.
Gap junctions within these regions provide low resistance pathways for current flow
between adjacent cells~\cite{Klab2012a}.
Each myocyte is connected longitudinally to at least two and laterally to
at least one neighbouring cell~\cite{spach1995stochastic,stinstra2010incorporating}.
The cell membrane forms invaginations, called transverse tubuli,
that penetrate into the interior of myocytes
to facilitate the rapid transmission of the action potential into the cell
and to synchronize \textcolor{review1}{calcium} release throughout the cell to enhance contractile forces \textcolor{review1}{\cite{Pagel2018}}.

Myocytes require a large number of mitochondria
to generate the metabolic energy needed for contraction and relaxation.
%The contractile elements in myocytes, referred to as sarcomeres,
%contain cytoplasm, the mitochondria and other intracellular organelles.
Sarcomeres are the fundamental contractile units of a myocyte
that are bounded by the Z-line.
A sarcomere consist of thick myosin filaments (forming the dark A-Bands)
which slide along thin actin filaments (forming the light I-bands).
The actin filaments are anchored to the Z-line and project about $1~\mu m$
toward the center of the sarcomere.
The thick filaments are about $1.6~\mu m$ in length and are centered around the M-line
where thick filaments are connected with each other by radial crosslinks.
The maximum force that can be generated by a sarcomere at any sarcomere length
is governed  by the degree of overlap between thin and thick filaments.
Optimal overlap is assumed at a sarcomere length of $2.0 - 2.2~\mu m$.
Thus stretching beyond the optimal length reduces maximum force down to zero
at a length of \SI{3.6}{\um} where no overlap is possible anymore.
At shorter sarcomere lengths during contraction maximum force also reduces
as a crossover between thin filaments occurs,
which impedes effective formation of crossbridges.
Under normal conditions sarcomeres operate along the ascending limb
of the length-tension relation in the range \SIrange{1.65}{2.3}{\um},
which is one mechanism contributing to the Frank-Starling effect~\cite{Frank1895a},
that is, the increase in force with preload.
%The length of an individual sarcomere is approximately \SI{2}{\um}
%and dependent on the state of contraction.
The contractile proteins are surrounded by a sarcoplasmic reticulum network,
which is the primary calcium ion reservoir \textcolor{review1}{\citet{Pagel2018}}. \autoref{fig:Fig3}, (b) shows a detailed schematic
of the myocyte's ultrastructure.

In contrast to other cell types in the human body, the turnover of adult myocytes
is very slow, at a rate of approximately \SIrange{0.5}{1.0}{\%} per year \cite{Koit2012a},
and myocytes do not multiply by cell division.
At birth, there are around \num{6} billion myocytes available in a healthy
human \textcolor{review1}{heart} and this number does not change significantly
in a \textcolor{review1}{lifetime} \cite{Leri2015a}.
Compared to other constituents, this occurs at \textcolor{review1}{a} rate slow enough
to assume that there is essentially no turnover in the cells themselves \textcolor{review1}{\cite{Robb2005a}}.
However, myocytes can grow in size due to sarcomerogenesis,
where new sarcomere units are created and deposited \cite{Robb2005a}.
In the healthy myocardium, around \num{50} sarcomeres make up one myofbril,
and \numrange{50}{100} parallel myofibrils result in a myocyte.
%Sarcomeres have an optimal length in which they operate most efficiently,
%known as the Frank-Starling mechanism~\cite{Frank1895a}.
%Deviation from this optimal length leads to a significant loss of function.

Myocytes are sensible to their environment. Mechanical stimuli are transmitted between
the \textcolor{review1}{cytoskeleton} and the ECM via transmembrane integrin proteins.
\citet{Bray2008a} showed that a modification in the ECM surrounding myocytes
results in a dynamic reconfiguration of the shape and intracellular architecture.
They cultured myocytes constrained to different shapes and concluded,
that the myofibrillar arrangement in square myocytes
(corresponding more to myocytes from hearts undergoing concentric hypertrophy) was almost isotropic,
whereas for myocytes with a width-to-length ratio of $1:7$
(corresponding to a more healthy myocyte ratio of \num{7.5}) the myofibrils were arranged
along the long axis, as shown in \autoref{fig:Fig8}.
\autoref{fig:Fig8} shows
that for myocytes with a higher length-to-width ratio the focal adhesions (seen in panels ii)
were larger in size compared to square samples. Focal adhesions are structures which
conduct mechanical and biochemical stimuli from the ECM to the cell
and through which a cell connects to the surrounding ECM\@.

Individual myocytes are surrounded and interconnected by endomysial collagen.
These are further connected by perimysial collagen, forming sheets, which are
connected via wavy perimysial collagen fibers. This is described in detail
in~\autoref{sec:collagen}.

In the healthy myocardium, myocytes are arranged along \textcolor{review1}{a} defined orientation, see \autoref{fig:Fig3}.
At the LV epicardium the prevailing orientation of myocytes, referred to as \lq fiber orientation\rq,
is around \SI{-60}{\degree} relative to the circumferential equatorial direction
and at the endocardium around \SI{90}{\degree} \cite{Leonard2012}.
\citet{Karl2000a} have fitted a distribution function to measure myocyte and
eart and observed a relatively small
dispersion in myocyte orientation of \SIrange{12}{15}{\degree},
which got exacerbated with disease progression.

%%%%%%%%%%%%%%%%%%%%%%%%%%%%%%%%%%%%%%%%%%%%%%%%%%%%%%%%%%%%%%%%%%%%%%%%%%%%%%%
\subsubsection{Collagen}\label{sec:collagen}
The ECM provides a scaffold defining the geometry and muscular architecture
of the heart and is hence responsible for its passive mechanical properties.
A detailed knowledge of the structure in health and disease is hence necessary
to establish meaningful computational models of G\&R.

Collagen, which is a significant part of the ECM, is the main load-bearing
structure in the myocardium, although the collagen content in the heart is
relatively low with around \SIrange{2}{4}{\%}. Despite the small fraction its
remodeling has significant effects of the mechanical behavior throughout
the whole cardiac cycle~\cite{Wang2016a}.

Collagen type I and III are the fibrillar collagen types of the
myocardium \cite{Weber1989}. Type I collagen accounts for approximately
$85\%$ of the collagen in the heart and consists of cord-like, parallel
fibrils, whereas type III collagen consists of a fine network of fibrils
and accounts for approximately \SI{15}{\%} of the total collagen
mass \cite{Leonard2012}.
Collagen \textcolor{review1}{has been reported to be degraded and replaced in the healthy heart at a rate of about \SI{0.6}{\%} per day up to \SIrange{1}{6}{\%} per day \cite{Macchiarelli2002}}
and a half life of \num{80}~to~\num{120}~days~\cite{Weber1989b}.
Initially the classification \textcolor{review1}{into perimysial and endomysial }used for cardiac collagen was defined for skeletal
muscle and only afterwards applied to the heart \cite{Caul1979a} to describe
different levels: Epimysial collagen denotes the layer of connective tissue
surrounding the myocardium. Perimysial collagen, which consists mainly of
collagen type I, envelops and connects groups of myocytes, forming so called
sheets~\cite{LeGrice1995a,Pope2008a,Leonard2012}.
Its fibers are often coiled and capable of storing energy,
which can be released during contraction or relaxation
at volumes larger or smaller than the unpressurized volume, respectively.
In a normal LV the end-systolic volume is below the unpressurized volume,
thus release of stored energy promotes re-lengthening of myocytes
during early diastolic filling~\cite{Weber2013},
which is referred to as elastic recoil.
Endomysial collagen, mainly being formed by collagen type III, connects and
surrounds individual myocytes and capillaries. In systole, most of the wall
stress is born by myocytes,
while endomysial collagen transmits force and keeps myocytes aligned.
During \textcolor{review1}{diastolic filling}, perimysial collagen uncoils and straightens,
eventually resisting further extension, which accounts for the steep part of
the end-diastolic-pressure-volume-relation,
thus protecting myocytes from becoming overstretched \cite{Fomo2010a}.
Perimysial collagen may play a key role in cardiac performance,
as it interconnects all contractile elements in the myocardium~\cite{Fedak2005}.

\citet{Macchiarelli2002} examined the microstructure of healthy endomysial
collagen of rabbit hearts using a method which allowed the isolation of
collagen by removing other ECM components as well as cells. They were able
to show that individual myocytes are connected by laminae, which envelop
the whole surface of vessels and myocytes, see~\autoref{fig:Fig4}, (a) and (b).
They hypothesized that this structure might provide a protection against
overstretch of individual myocytes.

Perimysial collagen can be further subdivided into three groups:
(1) a web-like structure surrounding muscle layers,
(2) long cords which are aligned with the fiber axis, and
(3) cords connecting sheets to each other and hence crossing cleavage
planes in a perpendicular manner relative to the long axis of
myocytes \cite{Leonard2012}.
\citet{Robinson1987} examined hearts of rats and hamsters and reported the
lateral insertion of perimysial collagen struts into the Z band of myocytes,
see~\autoref{fig:Fig4}, (c) and (d).
This organization exhibiting tight coupling inside sheets but loose coupling
between these provides the potential of significant deformation between the
sheets. In the healthy heart, the turnover rate of collagen \textcolor{review1}{has been reported to lie around \SI{0.6}{\%} per day~\cite{Weber1989b}  to \SIrange{1}{6}{\%} per day \cite{Macchiarelli2002}} is
, which is low in comparison to non-collageneous
protein turnover (\SI{7.2}{\%} per day)~\cite{Weber1989b}.
%
%%%%%%%%%%%%%%%%%%%%%%%%%%%%%%%%%%%%%%%%%%%%%%%%%%%%%%%%%%%%%%%%%%%%%%%%%%%%%%%
\subsubsection{Fibroblasts}
Fibroblasts are the predominating cells in the myocardium in terms of numbers,
making up, together with the other interstitial cells (macrophages, endothelial cells,
smooth muscle cells), over $75~\%$ of all myocardial cells.
They synthesize and regulate the degradation of the ECM and hence maintain
the structural integrity of the myocardium~\cite{Fedak2005}.

Fibroblasts originate from two sources. Firstly they differentiate from
epicardial-derived-cells directly in the heart which originate from events
taking place during embryonic development.
These cells can alternate between fibroblasts and a $\alpha$-smooth muscle actin expressing myofibroblast
phenotype. Eventually most epicardial-derived-cells assume the fibroblast
phenotype. However, during certain pathologies fibroblasts can differentiate into
myofibroblasts and have a significant influence on the mechanics of the
remodeling myocardium, see~\autoref{sec:change}. Additionally, fibroblasts
have been  recently reported to originate from progenitor stem cells from
the circulation \cite{Porter2009a}.

\emph{In vitro} it has been shown that the arrangement of fibroblasts is highly hierarchical, being arranged in sheets
and strands running parallel to muscle fibers and hence being able to
maintain the structural integrity within the myocardium. \emph{In vivo} this arrangement is less clear and
has been controverially discussed \cite{ongstad2016fibroblast}.
Fibroblasts do not only couple electrically between themselves,
but also hetero-cellular coupling between fibroblasts and myocytes has been observed \cite{Porter2009a}.
While this is well established \emph{in vitro},
to which extent this is the case \emph{in vivo}
and whether the coupling is functional and relevant is less clear~\cite{kohl14:_native,ongstad2016fibroblast}.
In addition to their role in maintaining the integrity of the ECM,
%\todo{communication can't be electronic as there are no electrons involved!}
fibroblasts \lq communicate\rq ~with myocytes through mechano-electric and electrophysiological signals
and are hence able to remodel the myocardium responding to
altered states of mechanical loading~\cite{VanPutten2016}. Fibrosis has been
viewed as an insulator to electrical conductivity until recently. However,
recent studies suggest that fibroblasts, although electrically unexcitable,
are able to contribute to the electrophysiology of the myocardium. In cell cultures,
fibroblasts have been reported to contribute to \textcolor{review1}{synchronizing} the contraction
between individual myocytes \cite{Gosh1969a}. More recently, \citet{Kamkin2003a}
have shown that fibroblasts can alter electrophysiological properties of
myocytes in coculture. The transfer of electrical signals occurs through
gap-junctions which have been shown to form between myocytes and fibroblasts
in co-culture \emph{in vitro} \cite{Gaude2003a}.

Fibroblasts can sense mechanical stimuli through their cytoskeleton,
integrins and stretch-activated channels~\cite{Creemers2011} (see~\autoref{fig:Fig6},
(a) for an electron micrograph of the complex cytoskeletal network) and respond to it
via changes in their resting potential. The ability to respond to a variety
of mechanical, electrical and chemical stimuli is pivotal to maintain normal
cardiac function and a unique property compared to fibroblasts from other tissues.
Fibroblasts do this by expressing proteolytic enzymes such as matrix
metalloproteinases (MMPs) and growth factors, synthesizing collagen,
differentiation into the myofibroblast phenotype (see~\autoref{sec:change}),
proliferation and migration, see~\autoref{fig:Fig6},~(b).
In organs such as the skin, fibroblasts are protected against severe mechanical
stimulation, which is one of the main differences to cardiac fibroblasts, which
are subjected  to cyclic mechanical loads and are able to sense changes to the
physiological cyclic loading.~\cite{Porter2009a}
%
%%%%%%%%%%%%%%%%%%%%%%%%%%%%%%%%%%%%%%%%%%%%%%%%%%%%%%%%%%%%%%%%%%%%%%%%%%%%%%%
\subsubsection{Elastin and Proteoglycans}
In a recent review by \citet{Fomo2010a} the authors were surprised to find
that despite hints that elastin and proteoglycans may play important roles
in the mechanical behavior of the heart, their contribution has received
little attention. Additionally, the contribution of all these components
is much better understood in other soft tissues such as arteries than in
the myocardium.

As the heart muscle exhibits a limited elasticity and as there is less elastin
present in the myocardium than in the vasculature the role of myocardial
elastin it not yet understood. There have been even less studies on myocardial
proteoglycans, although they may play important roles in the lubrication of
the myocardial space for contraction and sequestering bioactive proteins such
as growth factors for cellular signaling~\cite{Fomo2010a}.

No data are to be found on the change of these constituents during pressure overload,
although they are known to play significant roles in other diseases of the vasculature.
This may be an interesting starting point for future research.
%
%%%%%%%%%%%%%%%%%%%%%%%%%%%%%%%%%%%%%%%%%%%%%%%%%%%%%%%%%%%%%%%%%%%%%%%%%%%%%%%
\subsection{Changes in the Myocardial Structure during Pressure Overload}%
\label{sec:change}
Altered circumstances such as pressure overload caused by aortic stenosis were
reported to trigger significant and complicated remodeling processes in the heart.
In the 1980s a series of studies and reviews by
Weber~et~al. (e.g.~\cite{Weber1987, Weber1989}) shed some light on the underlying mechanisms.
After experimental induction of pressure overload,
the synthesis of non-collageneous proteins increased by $75\%$,
and returned to control levels after two weeks \textcolor{review1}{while pressure overload was experimentally maintained}. Myosin synthesis
was found to be increased right on the onset of pressure overload, which
leads to myocyte hypertrophy. The response of collagen synthesis was much
slower but persisted for a longer period of time. A six- to eight-fold
increase to $4\%$ per day was observed and remained increased \textcolor{review1}{threefold} at
examination at two and four weeks. As fibroblast proliferation was only seen
at a later stage, \citet{Weber1989b} concluded that collagen
synthesis is first carried out by existing fibroblasts, and fibroblast
proliferation only follows thereafter. This finding was later reproduced,
e.g., in 2002 by \citet{Akdemir2002}.
At the onset of pressure overload, more collagen type III was found to be
deposited, which expresses in thinner fibers~\cite{Akdemir2002}.
Interestingly, only ventricles directly affected by pressure overload showed
an increase in collagen content --- this finding excluded the possibility that
collagen synthesis was triggered by circulating growth factors, which would
trigger collagen growth in the whole heart \textcolor{review2}{\cite{Akdemir2002}}.
During the onset of pressure overload, collagen degradation was observed to increase,
but returned to base levels soon after \textcolor{review2}{\cite{Bonnin1981}}.
As myosin synthesis was reported to occur more rapidly
at the onset of pressure overload, myocyte hypertrophy and a decline in
collagen content can be seen at the early stage of pressure overload~\cite{Bonnin1981}.
The increase in myocyte diameter is accompanied by little or no change in
myocyte length \textcolor{review2}{\cite{Kajstura2004}}. The lateral expansion of pre-existing myocytes is considered
to be the only cellular process to increase the wall width to normalize peak
systolic wall stresses \cite{Kajstura2004}. Additionally, the area of the transverse
tubuli in myocytes was shown to be elevated \cite{Ferr2002a}.
Myocyte growth was reported to stabilize soon~\textcolor{review2}{\cite{Weber1989b}}.
However, collagen synthesis persisted and exceeded its degradation, increasing collagen
content significantly above baseline values after four
weeks or more~\cite{Weber1989b}.

These processes are the beginning of a continuous structural remodeling,
resulting in an increased dimension of collagen fibers seen at later stages
of hypertrophy, expanded intermuscular spaces being filled with perimysial
collagen and an increase in the fibrous mesh-work surrounding
myocytes~\textcolor{review2}{\cite{Akdemir2002}}.
Additionally, collagen accumulates around small intra-myocardial coronary
arteries~\textcolor{review2}{\cite{Akdemir2002}}.
Fibrosis without cell necrosis was termed reactive fibrosis and is dominant
in earlier stages of hypertrophy due to pressure overload~\textcolor{review2}{\cite{Weber1988}}.
In the beginning, this adaptive process enhances systolic stiffness but also
increases diastolic stiffness. When pressure overload persists, those
initially positive, adaptive changes lead to cell necrosis and reparative fibrosis.
When cell death occurs and mechanical overload persists, fibroblasts can
phenotypically transform into myofibroblasts and express $\alpha$-smooth
muscle actin myofilaments~\textcolor{review2}{\cite{VanPutten2016}}.
In the beginning, fibroblasts do not yet contain
stress fibers~\textcolor{review2}{\cite{VanPutten2016}}. They form from cytoplasmic
actins and myosins responding to \textcolor{review1}{a} mechanical stimulus and produce a ``proto-myofibroblast''~\textcolor{review2}{\cite{VanPutten2016}}. These stress fibers
end in transmembrane integrin-containing adhesion complexes which are attached
to the surrounding ECM and can transmit and also sense mechanical
loads~\textcolor{review2}{\cite{VanPutten2016}}.
When the elevated mechanical load persists, proto-myofibroblasts fully develop
into myofibroblasts with $\alpha$-smooth muscle actin being incorporated into
the stress fibers~\textcolor{review2}{\cite{VanPutten2016}}. At this stage
myofibroblasts can generate force~\cite{VanPutten2016},
anchoring to the ECM and contracting, thereby closing spaces vacant after myocyte
death \cite{Gabb2003a}. In combination with inflammatory cells, which have a
short lived presence at the site of injury, myofibroblasts are responsible for
collagen fibrillogenesis and scar genesis by increasing the secretion of
ECM-degrading metalloproteinases (MMPs) and collagen turnover~\textcolor{review2}{\cite{VanPutten2016}}.
Additionally, \citet{Weber2013} reported an adverse
cell-to-cell interaction between myofibroblasts and cardiomyocytes,
negatively influencing the electrical behavior of the myocardium.
Lost cardiomyocytes are replaced by stiff, cross-linked collagen type I,
called reparative fibrosis~\textcolor{review2}{\cite{Weber2013}}.
Additionally, fibroblasts invading from the
adventitia of intramural coronary arteries and arterioles can trigger
reactive fibrosis~\textcolor{review2}{\cite{Weber2013}}. In this case, they again
differentiate into myofibroblasts and ensnare cardiomyocytes with a weave of fibrous tissue,
causing their atrophy~\textcolor{review2}{\cite{Weber2013}}.  As a consequence the occurrence of fibrosis is assumed to be implicated in the transition from compensated
hypertrophy to systolic pump failure~\cite{Weber2013}.
Eventually this manifests as diastolic failure,
where EF is preserved but the capability to increase cardiac output when necessary is
impaired~\cite{LeGrice2012b}. Hypertrophy is accompanied by a significant
decrease in capillary density, hence the supply with blood decreases and
further promotes myocyte death~\cite{Hein2003a}.
At this stage, collagen type I represents more than \SI{90}{\%} of all collagen,
bridging the gaps which originate from cell loss.
The increasingly dense network of collagen fibers surrounding myocytes
restricts the stretching of muscle fibers during diastole and hence reduces
the length-dependent ability to generate force during systole.
These changes lead to myocyte atrophy \cite{Weber1989b}.
A further increase in collagen synthesis is then triggered, as myocardial
injury triggers the proliferation of fibroblasts to myofibroblasts
and influx of inflammatory cells, which in turn secrete matrix
metalloproteinases (MMPs) \cite{Leonard2012}.
This is characterized by an ongoing dilatation of the LV and
a reduction in EF\@.

\citet{Hein2003a} have proposed a detailed schematic of continuous
remodeling in pressure overload after carefully examining myocyte death,
fibrosis percentage and other factors in three different groups of patients
with aortic stenosis, categorized according to their EF\@.
Based on these findings they proposed a schematic of continuous remodeling in pressure overload
induced hypertrophy. \autoref{fig:Fig7} shows an adapted version including
some findings from this review. \autoref{tab:pathogenesisTable} summarizes
the main events of the pathogenesis following pressure overload which are described
in this chapter.

For a comparison of healthy myocardium and myocardium affected by pressure
overload, see~\autoref{fig:Fig5}. Many experimental and clinical studies have
been published in the last $30$ years considering pressure overload.
However, most of them do not provide any quantification of the data, making
them less suitable for computational modeling.
\autoref{tab:expTab} gives an overview \textcolor{review1}{of} quantified findings of studies
which focused on biopsies from the LV free wall during AVR\@.
As can be seen, absolute values for instance for myocyte diameter vary
slightly over a rather narrow range, but others, such as volume fraction of
collagen, can vary significantly between studies. For example, the study
by~\cite{Hess1981a} found a collagen volume fraction of \SI{2}{\%} in healthy
myocardium (similar to several other studies),
whereas for example~\cite{Hein2003a} reported a value of \SI{12}{\%}.

\autoref{tab:expTab} additionally includes studies, which grouped patients
with aortic sclerosis according to their EF,
following the hypothesis that a reduction in EF correlates with
the severity of the disease. Other groups, such as,
e.g.,~\cite{Krayenbuehl1989} or more recently~\cite{Treibel2018} performed
subsequent biopsies at different time points pre- and post AVR
to study the ongoing remodeling, which makes these studies
particularly interesting as an input for computational studies.
Additionally, clinical characteristics such as heart rate (HR), EF,
end diastolic and systolic pressure (EDP and ESP, respectively)
and LV wall thickness were included into the table as they may proof useful
for the validation of computational studies.
\subsection{Mechanotransduction/Mechanosensing}\label{sec:mechanosensing}
%%%%%%
Cells such as myocytes or fibroblasts are able to sense and respond to mechanical
stimuli, called mechanosensing, and translate them into intracellular biochemical
stimuli, called mechanotransduction. In the heart, increased mechanical stimuli
lead to upregulated protein synthesis and myocyte hypertrophy.~\cite{Weber2013}

Over at least the past \num{20}~years there has been a debate on what type of
mechanical signals cells sense. In \num{2001}, \citet{Hump2001a} postulated that cells cannot respond directly to mathematical concepts
such as stress, strain, or strain-energy. They rather sense local pushes and
pulls due to conformational changes in the environment and
respond to these. However, the mathematical concepts of, e.g., stress
and strain still remain an invaluable tool to identify empirical correlations
and model these.
As understanding the fundamental mechanisms of cell response
to mechanical stimuli is essential for formulating constitutive equations
we give a summary on how cells in the myocardium sense their environment and how
they response to mechanical stimuli.

\citet{Bray2008a} stated that the contractile architecture of a myocyte is highly
sensible to changes to the environment and even to intracellular boundary
conditions or the positioning of the nucleus relative to the surroundings.
As the ECM provides the surroundings to individual myocytes, pathological
changes as they occur during pressure overload are clearly changing the myocyte's
boundary conditions and hence contribute significantly to intracellular remodeling.

All eucaryotic cells, such as myocytes or fibroblasts, have a cytoskeleton made of
actin, which readily remodels to adjust cell shape. When sarcomeres develop, focal
adhesions form at cell membranes and anchor the cytoskeletal elements to the ECM\@.
Through these focal adhesions actin filaments are able to emanate outside. In the
myocyte, the connections to the surrounding ECM and hence the transmission of force
form the contracting myocyte to the ECM are called costameres. In the course of
sarcomere development pre-myofibrils are formed close to the membrane which look
like short sarcomeres between Z-bodies. With time they grow apart and develop longer
A-bands. Then sarcomeres are assembled with template proteins such as titin, nebulin,
and myosin binding protein C and actin capping proteins define the thin filament length~\cite{Russell2010}\textcolor{review1}{.}

Although the process of sarcomere formation is well understood,
it remains still unknown how actin filament insertion during new sarcomere addition in the adult myocyte takes place.
Adding to the complexity of this problem,
sarcomeres can be added in-parallel or in-series, where the processes are not fully understood yet either
\cite{Russell2010}.
In 1990, \citet{Dix1990a} studied myofibrillogenesis in stretched
muscle fibers from adult skeletal muscles. They found that sarcomeres are added at
the end of fibers near myotendon junctions to accomplish lengthening. The analogous
structure in myocytes is the intercalated disc, where \citet{Perr2003a} found a
denser architecture in myopathic hearts compared to healthy ones, suggesting a
possible site where sarcomeres might be added. Other possible locations for sarcomere
addition are the middle of the muscle or the laterally oriented Z-disc, which is a
lateral extension of the focal adhesion. Although no study has been able yet to confirm

translated there for in-series or in-parallel addition of thin filaments.

Mechanical changes in the cell's environment can be detected by the above mentioned
focal adhesion complex. The mechanical deformation of integrins is most likely the
first step in a complex intracellular signaling cascade leading to cytoskeletal
rearrangements \cite{Creemers2011,Russell2010}.
For details on the cellular signaling pathways the interested reader
is referred to, e.g.,~\citet{Russell2010}.
Integrins are trans-membrane proteins, linking across the cell membrane from
intracellular actin filaments to specific components in the ECM\@. They can mediate
intracellular signals in both ways --- from the ECM to the cell and vice versa. Integrins
can mediate cell contraction which in turn activates latent TGF-$\beta$ stored
in the ECM\@. TGF-$\beta$ is a regulator of fibrotic processes throughout the body.
The activation takes place through the transmission of intracellular force via integrins.
The stiffness of the surrounding ECM has been shown to influence the ability to activate
latent TGF-$\beta$ directly, as an increased ECM stiffness results in a higher
resistance to cell forces. The release of TGF-$\beta$ has been shown to be higher on
stiff substrates compared to compliant ones \cite{VanPutten2016}.

In myocytes, another mechanotransduction site is the Z-disc. Here, thin actin
filaments insert and reverse the polarity. Additionally, perimysial type I collagen
inserts laterally to the Z band \cite{Robinson1987}. The longest protein known,
\textcolor{review1}{titin}, interacts with multiple proteins within the Z-disc and has elastic sequences in
the I-Band. These elastic structures serve as springs which store strain-energy during
diastolic filling and recoil during systole. The mechanical load is sensed by this complex and
activates signals through thin and intermediate filaments which anchor to the focal
adhesion complex, see \autoref{fig:Fig9}~\cite{Knoe2002a,Mill2004a}.
\citet{Weber1989} and~\citet{Knoe2002a} hypothesized that this linkage
might translate elevated mechanical stimuli directly into fibroblast
proliferation and sarcomere addition.
\citet{Knoe2002a} have been able to show that the titin/Z-disc structure plays an
essential role in a whole complex triggering pathways following mechanical stimuli.
Additionally to this purely mechanical trigger, mechanically induced release of peptides from
neighbouring cells such as fibroblasts or endothelial cells serve as hormonal stimuli
for triggering the above mentioned pathway in the Z-disc. For a detailed description
of the pathways the interested reader is referred to~\cite{Knoe2002a}.

Fibroblasts do not contain stress fibers. However, if they are subjected to a lasting
mechanical stimulation they can change their phenotype to myofibroblasts
(see~\autoref*{sec:change}) which have stress fibers. There, intracellular stress-fibers terminate in the
adhesion complexes, connecting myofibroblasts to their surrounding ECM and making
it possible to transmit and perceive mechanical stimuli \cite{VanPutten2016}.
Integrins link the contractile actin-myosin cytoskeleton to the ECM \cite{Wang1993a}.
%\todo{In myofibroblasts? These do have a contractile machinery? yep. im Phenotypen des myofibroblasten
%haben sie einen ähnlichen kontraktilen Mechanismus wie SMC, um zB gesundes Gewebe um tote Zellen herum zusammen zu ziehen.}
In the case of fibroblasts, this linkage allows them to ``pull'' on the surroundings
and hence probe the mechanical state of the environment by matching the extracellular
resistance with intracellular force developed by actin/myosin contractility of their
stress fibers. Myofibroblasts get activated  after injuries such as myocardial
infarction where scarring occurs or during overload, where strain stiffening
occurs due to the stretched collagen fibers. In both scenarios the cells sense ECM
stiffening \cite{VanPutten2016}.

It can be concluded that both myocytes and fibroblast exhibit mechanosensing
capabilities to react to relative changes to the homeostatic ranges of mechanical stimuli
they usually experience. These relative changes of mechanical stimuli
act as the triggers of complex signaling cascades, influencing cytoskeletal proteins \cite{VanPutten2016}.
%
%%%%%%%%%%%%%%%%%%%%%%%%%%%%%%%%%%%%%%%%%%%%%%%%%%%%%%%%%%%%%%%%%%%%%%%%%%%%%%%
\section{Discussion}
This review summarizes the recent developments in computational models of cardiac
G\&R, providing extensive summary tables on
details of the underlying modeling assumptions.
The number of attempts to build comprehensive G\&R models for the myocardium
under pressure overload is limited, suggesting that there is a pressing need
for structure based models of growth and remodeling, as many factors of the
pathogenesis leading eventually to heart failure (or to complete or incomplete
reverse remodeling in some cases) are not sufficiently well understood.

Microstructurally motivated computational models are able to provide a platform to ``explore'' with the complex contributions of multiple stimuli and, eventually, make it possible to establish models
which are able to provide clinically useful information.
Interdisciplinary investigations
are needed to tackle this problem, as driving factors are still only suspected but
not known, and a better knowledge of the complex interrelations is needed. Detailed
quantitative data is essential for model development and sound validation. In our view,
kinematic growth models are able to predict growth on a superficial,
phenomenological level, which might be enough for some applications. However,
to truly enhance the understanding of the basic pathogenesis and hence
to contribute towards identifying potential targets for medication or treatments,
we believe that the more detailed and microstructurally based constrained mixture models are likely to provide a more powerful framework.

To develop such models, however, more data are needed.
This reviews clearly highlights a paucity of data available
for characterizing pathological growth due to pressure overload
beyond a single time point. Quantitative data on, e.g., myocyte apoptosis,
fibroblast or myofibroblast cell density are scarce or not available at all.
Also, the quantitative data available on ECM volume fractions indicated rather
significant discrepancies between the studies.
These are most likely due to the different experimental protocols used
and also to inconsistencies in the definitions of collagen and ECM throughout the studies.
However, the data are still valuable as at least trends of development can be derived from the
quantitative data.

Finally, the question remains which trigger for growth computational models should use.
As discussed above, stress or strain are mathematical concepts and cannot be sensed by
cells. Nevertheless, they are useful metrics for the environment of continuum mechanics.
Hence, it should be clear, that these concepts are mere phenomenological auxiliary constructs.
To overcome this problem, \citet{Cyron2014b} introduced the concept of
mechanobiological stability, hypothesizing that aneurysm development might be viewed
as a mechanobiological instability. This concept might be more promising than looking
for mathematical, phenomenological triggers of growth such as stress or strain.

There are many challenges ahead to improve the current state of computational
G\&R models. The pathobiology of adaptation processes is highly complex, and
further complicated by the presence of various comorbidities
(hypertension, diabetes, renal disease, atrial fibrillation, vasculopathy, metabolic syndrome). \textcolor{review1}{As the active myocardial state changes due to G\&R, he link between active and passive mechanics should be modelled in future models as well.}
While the mechanistic links between organ-scale biomechanics and
biological factors at the cellular size scale remain poorly understood,
undoubtedly, a change in the mechanical stimuli experienced by a cell
can trigger complex changes within a cell, affecting cytoskeletal proteins and
ECM integrity (\autoref{sec:mechanosensing}).
There have been efforts to include, e.g., cell signaling~\cite{Ambr2019a}, \textcolor{review1}{the role of integrins using a "mechanical biodomain model" \cite{Sharma2018}}
or changes in gene expression or metabolism~\cite{Kim2018a}, but
microstructurally motivated, rigorously validated computational models of G\&R
are still in their infancy.

Addressing these various challenges to develop more advanced biologically sound computational G\&R models
that are able to quantitatively characterize the state of remodeling and allow predictions of the further disease progression
will require significant research efforts.
However, with the incidence of heart failure increasing dramatically in the ageing population,
the benefits of any advances in this direction appear highly rewarding.
%However, \emph{in-silico} models show high
%promise to be suitable for making clinical predictions on disease progression
%and for testing hypotheses on pathogenic mechanisms.
%%%%%%%%%%%%%%%%%%%%%%%%%%%%%%%%%%%%%%%%%%%%%%%%%%%%%%%%%%%%%%%%%%%%%%%%%%%%%%%
\section*{Conflict of Interest Statement}
The authors declare that the research was conducted in the absence of any
commercial or financial relationships that could be construed as a potential conflict of interest.
%
%%%%%%%%%%%%%%%%%%%%%%%%%%%%%%%%%%%%%%%%%%%%%%%%%%%%%%%%%%%%%%%%%%%%%%%%%%%%%%%
\section*{Author Contributions}
Developed the structure and arguments for the paper: JAN\@.
Performed the literature review: JAN\@. Wrote the first draft of the manuscript:
JAN\@. Contributed to the writing of the manuscript: JAN, CMA, GP\@. Wrote/designed
the abstracts: CMA, GP\@.
Agreed with manuscript results and conclusions: JAN, CMA, GP\@.
Made critical revisions: CMA, GP\@. All authors reviewed and approved of the final manuscript.
%
%%%%%%%%%%%%%%%%%%%%%%%%%%%%%%%%%%%%%%%%%%%%%%%%%%%%%%%%%%%%%%%%%%%%%%%%%%%%%%%
\section*{Funding}
The research was supported by the Grants F3210-N18 and I2760-B30
from the Austrian Science Fund (FWF), and a BioTechMed-Graz award
(Grant No. Flagship Project: ILearnHeart) to GP\@.
Additionally, this project has received funding from the European Union's
Horizon 2020 research and innovation programme under the
Marie Sk\l{}odowska--Curie Action H2020-MSCA-IF-2016 InsiliCardio,
GA No. 750835 to CMA\@.

\clearpage
%%%%%%%%%%%%%%%%%%%%%%%%%%%%%%%%%%%%%%%%%%%%%%%%%%%%%%%%%%%%%%%%%%%%%%%%%%%%%%%
% FIGURES
\section*{Figures}
%
%%% Please be aware that for original research articles we only permit a
%combined number of 15 figures and tables, one figure with multiple
%sub-figures will count as only one figure.
%%% Use this if adding the figures directly in the manuscript,
%if so, please remember to also upload the files when submitting your article
%%% There is no need for adding the file termination, as long as you indicate
% where the file is saved. In the examples below the files
\begin{figure}[ht]
  \centering
  \includegraphics{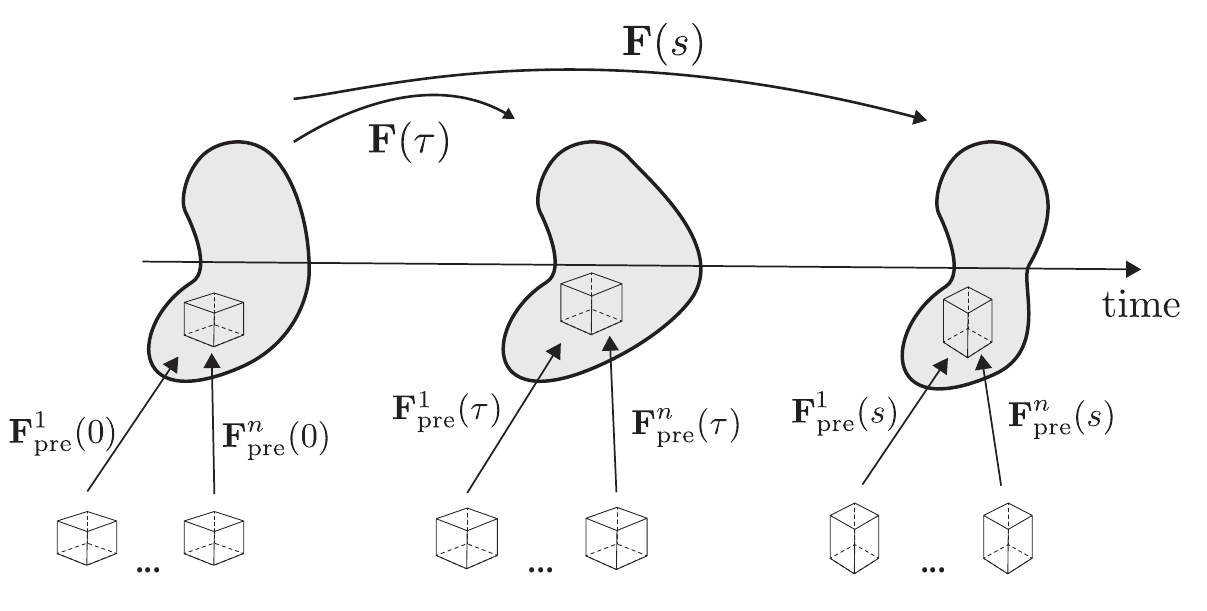}
  \caption{In constrained mixture models, a body is composed by $n$ individual
    constituents, each consisting of multiple mass increments which were deposited
    with a pre-stretch $\mathbf{F}_{\rm pre}(t)$ at different times.
    The elastic pre-stretch depends on the individual stress-free natural
    configuration of each constituent. All constituents undergo the same
    elastic deformation together, despite having been deposited with different
pre-stretches at different times. \textcolor{review2}{Adapted with permission from}~\citet{Cyron2017}.}%
  \label{fig:Fig2}
\end{figure}
\begin{figure}[ht]
  \centering
  \includegraphics[scale=0.8]{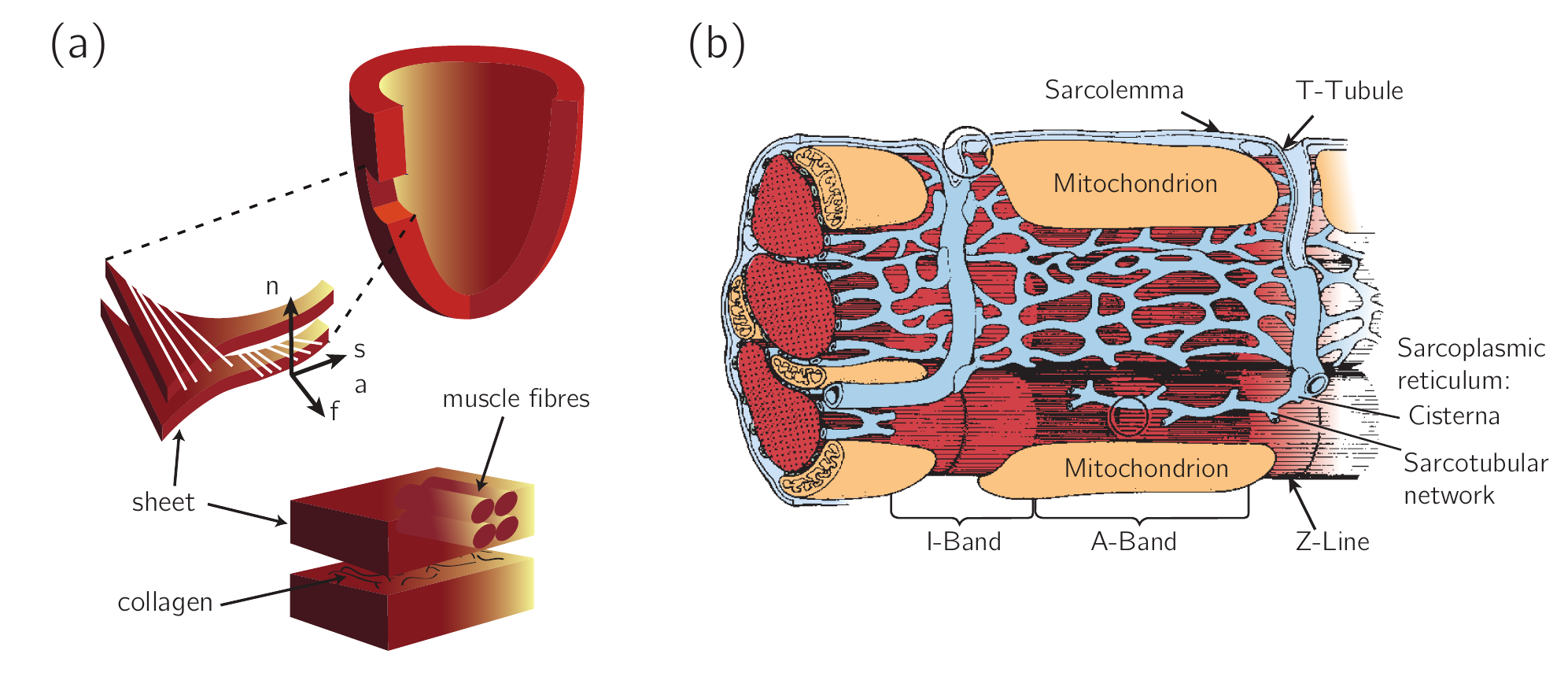}
  \caption{(a) Schematic drawing of the arrangement of myocytes,
    \textcolor{review1}{reproduced with permission from} \citet{Wang2016a}; (b) Schematic drawing of the structure
  of the myocyte, \textcolor{review1}{reproduced with permission from}~\citet{Pagel2018}.}%
  \label{fig:Fig3}
\end{figure}
\begin{figure}[ht]
  \centering
  \includegraphics[scale=0.8]{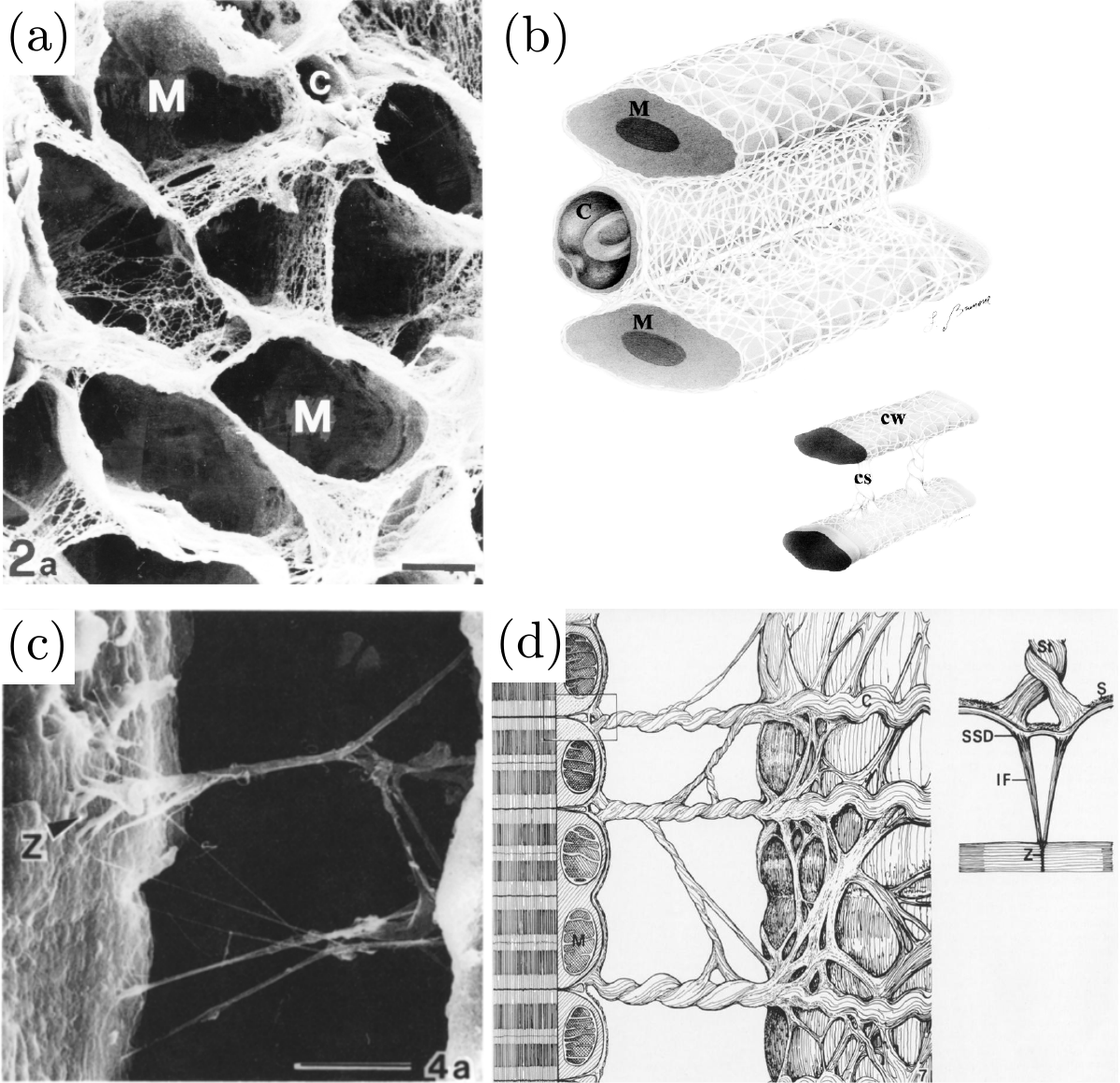}
  \caption{(a) Scanning electron micrographs showing endomysial collagen
      surrounding myocyte lacunae (M) and a lacunae of a capillary (C)
      surrounded by the same collagen, scale bar \SI{6.5}{\um}.
    (b) Schematic drawing of the distribution of endomysial collagen in a
      rabbit heart; a collagen weave (CW) enveloping myocytes (M) and
      capillaries (C) and collagen structs (CS) connecting single myocytes to
      each other.
    (a) and (b) \textcolor{review1}{modified with permission from} \citet{Macchiarelli2002}.
    (c) Scanning electron micrographs of struts of perimysial collagen anchored
      to the sarcolemma surface at the Z band plane (Z), scale bar \SI{10}{\um}, and
    (d) a schematic drawing of this structure. S:~sarcolemma, St:~strut,
      IF:~intermediate filament, M:~mitochondrion, SSD:~subsarcolemmal density,
      C:~collagen.
    (c) and (d) \textcolor{review1}{modified with permission from} \citet{Robinson1987}.}%
  \label{fig:Fig4}
\end{figure}

\begin{figure}[ht]
  \centering
  \includegraphics[scale=1]{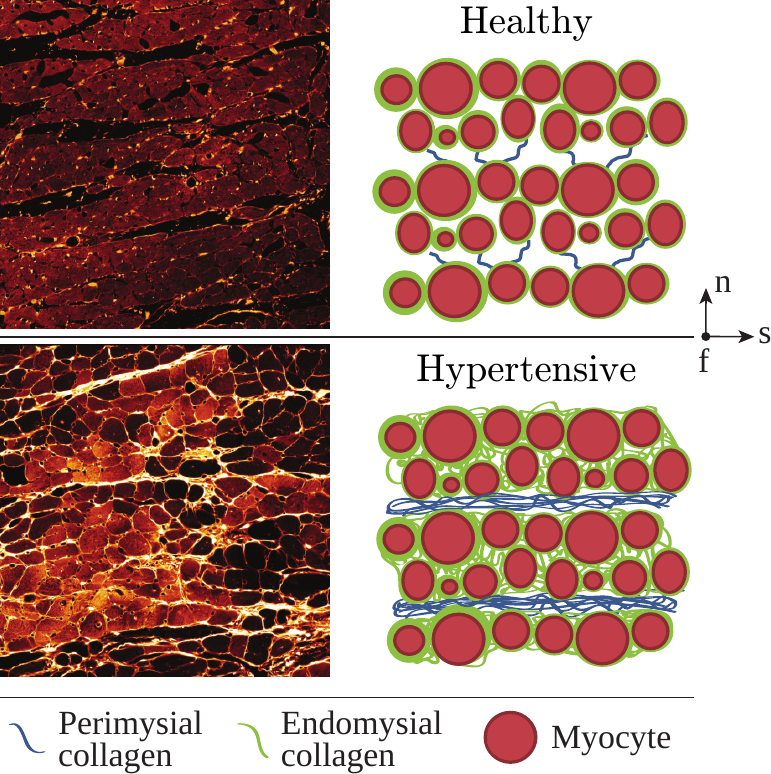}
  \caption{High resolution \emph{ex vivo} confocal images of tissue blocks
    from a healthy rat heart and schematics showing the arrangement of the
    constituents (top) and high resolution ex vivo confocal images showing a
    rat heart after remodeling due to pressure overload and the corresponding
    schematic (bottom), \textcolor{review1}{adapted with permission from}~\citet{Wang2016a}.}%
  \label{fig:Fig5}
\end{figure}

%\begin{landscape}
\begin{figure}[ht]
  \centering
  \includegraphics[scale=1]{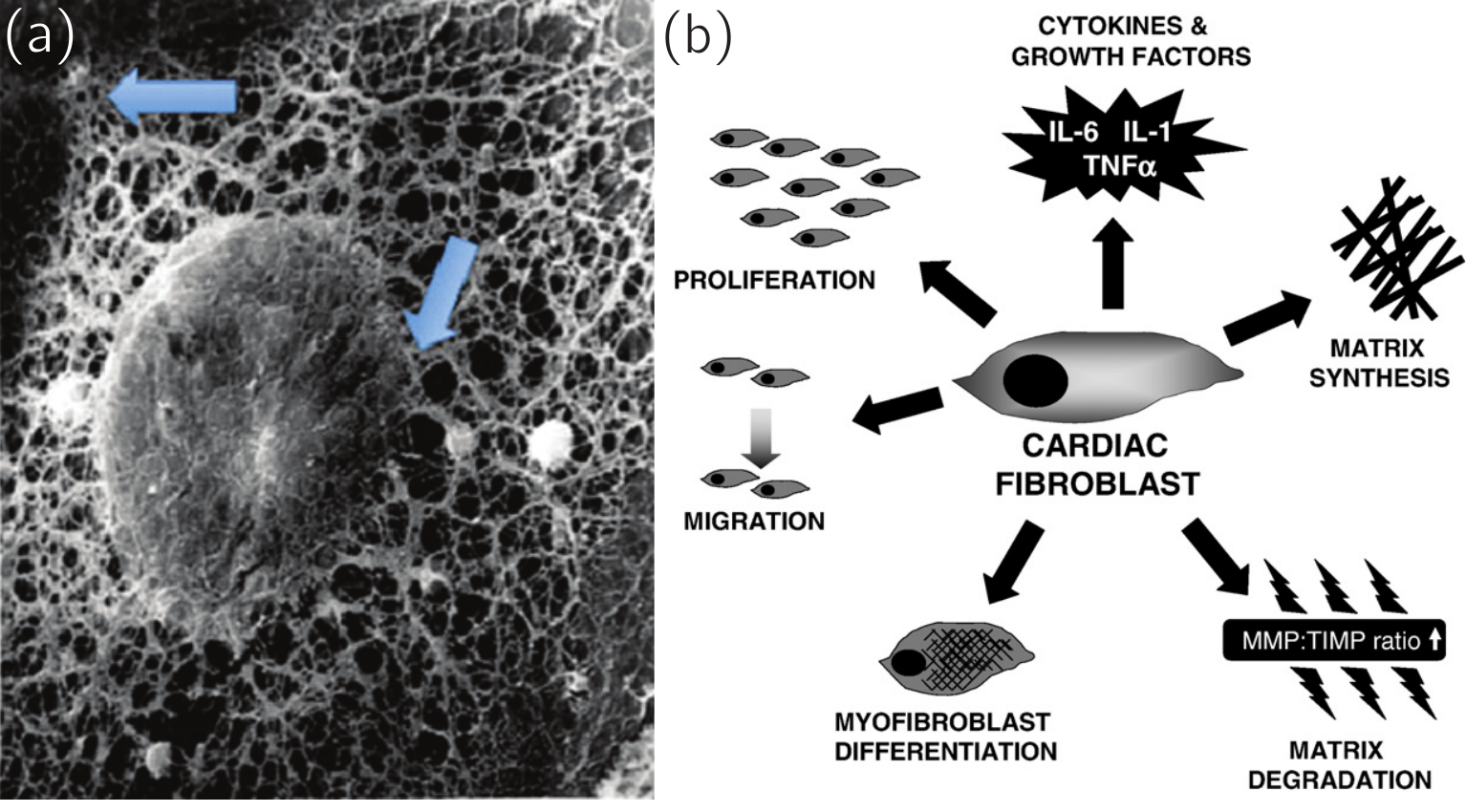}
  \caption{(a) A fibroblast with removed plasma showing the cytoskeletal network
    connecting an adhesion side on the plasma membrane with the connection
    to the nuclear membrane (arrows). \textcolor{review1}{Reproduced with permission from}~\cite{Borg2011}. (b) A schematic depicting the various responses of a cardiac fibroblast to environmental stimuli, including differentiation into another phenotype, migration, contribution to ECM turnover, secretion of growth factors and matrix degradation. \textcolor{review1}{Reproduced with permission from}~\cite{Porter2009a}. }%
  \label{fig:Fig6}
\end{figure}
%\end{landscape}

\begin{figure}[ht]
  \centering
  \includegraphics[scale=0.8]{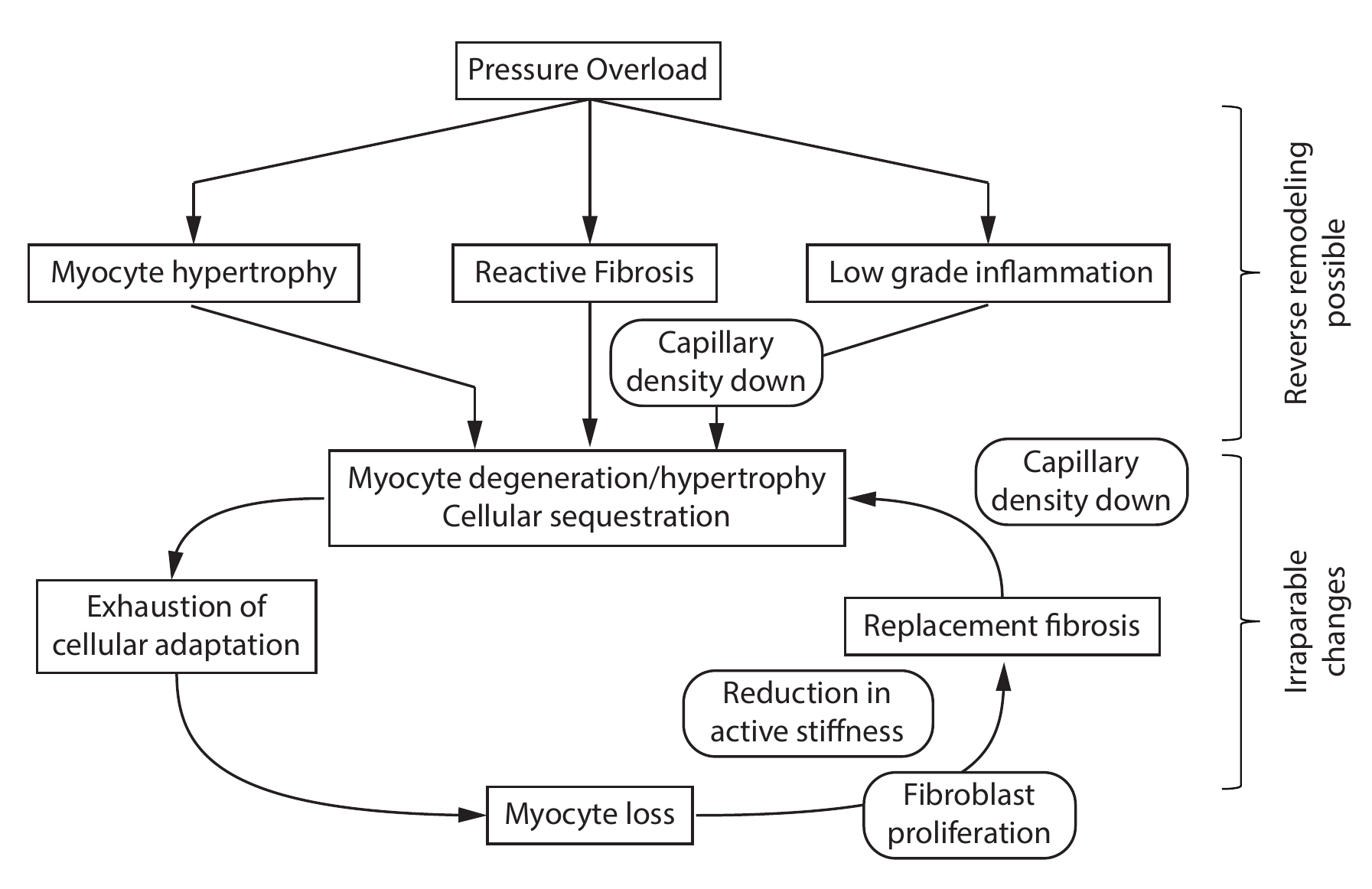}
  \caption{A schematic showing the pathogenesis of pressure
           overload induced hypertrophy. \textcolor{review2}{Based on}~\cite{Hein2003a}.}%
  \label{fig:Fig7}
\end{figure}
%
%\begin{landscape}
\begin{figure}[ht]
	\centering
	\includegraphics[scale=1]{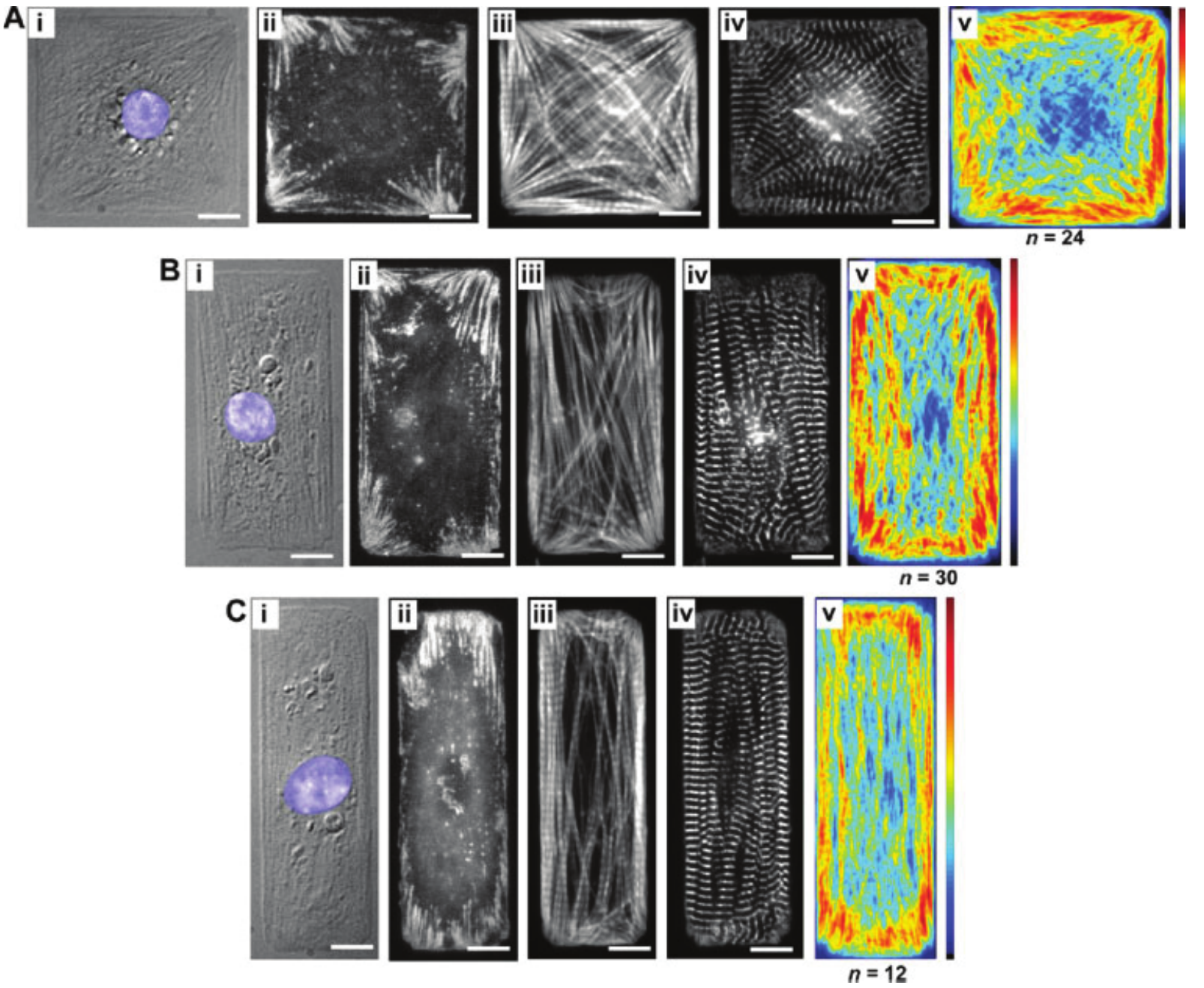}
	\caption{\citet{Bray2008a} printed ECM islands and placed myocytes
		on them to study the influence of the ECM on intracellular
		constituent alignment. The three cellular aspects are (A): 1:1, (B): 2:1 and (C): 3:1.
              (i) depicts a DIC image, (ii)--(iv) immunofluorescent stains for
vinculin (revealing focal adhesions), F-actin (staining I-bands) and sarcomeric $\alpha$-actin (revealing Z-bands). The average distribution of
F-actin is shown in (v).
\textcolor{review1}{Reproduced with permission from} \cite{Bray2008a}.}%
	\label{fig:Fig8}
\end{figure}
%\end{landscape}
%
\begin{figure}[ht]
	\centering
	\includegraphics[scale=1]{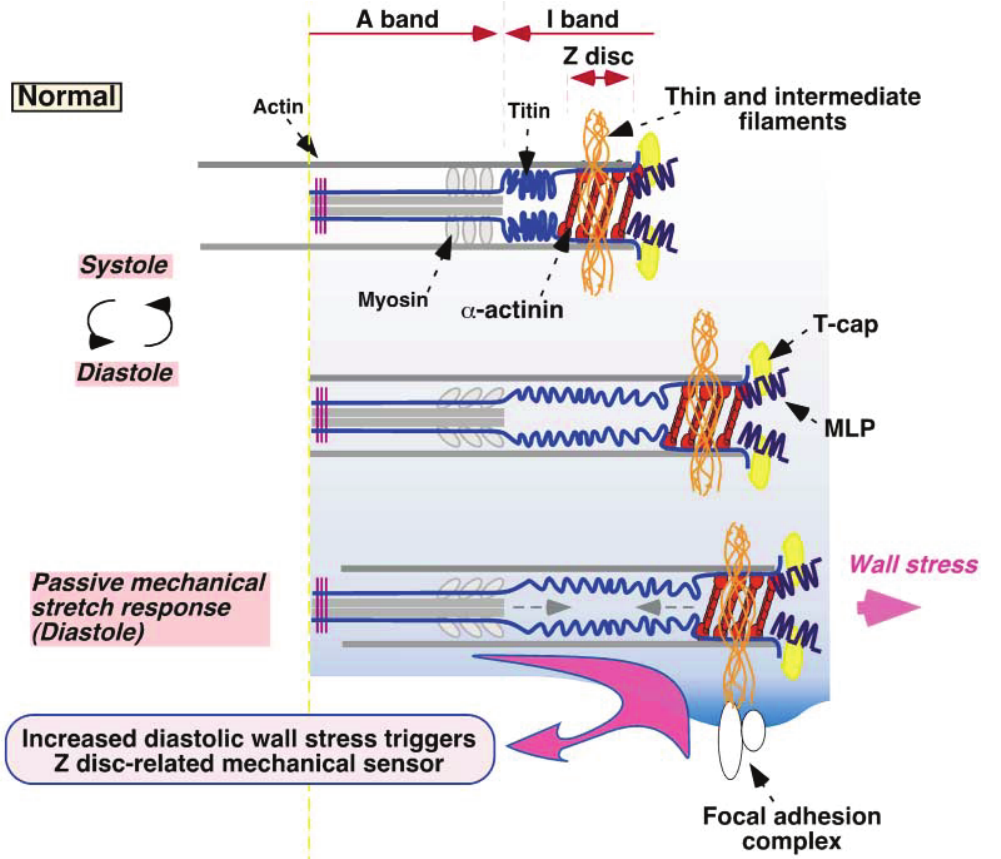}
	\caption{A schematic showing the interaction between \textcolor{review1}{thin} and intermediate
	filaments within the titin/Z-disc complex with focal adhesion complexes, hence
	serving as mechanosensors. Titin has elastic sequences in the I-band, serving as
	springs saving elastic energy during diastole and relasing it to regain the initial sarcomere length at systole. At peak diastole the titin elastic segments uncoil and add their
	contribution to ventricular wall distensibility. Increased stretch of the titin elastic segments is sensed and activates downstream signals for cardiac remodeling. \textcolor{review1}{Adapted with permission from}~\cite{Knoe2002a}.
}%
	\label{fig:Fig9}
\end{figure}

\newpage
%%%%%%%%%%%%%%%%%%%%%%%%%%%%%%%%%%%%%%%%%%%%%%%%%%%%%%%%%%%%%%%%%%%%%%%%%%%%%%%
% TABLES
\begin{landscape}
\begin{table}[htpb]
  \setlength{\extrarowheight}{1.5pt}
  \caption{{\small Overview of studies utilizing the kinematic growth
    theory to study G\&R in the heart.
    Note that studies on organogenesis, e.g., \cite{Lin1995,Ramasubramanian2006}
    are not included.
    $\mathbf{F}_{\rm{g}}$ is the inelastic growth deformation gradient;
    $\mathbf{M}_{\rm{e}}$ is the elastic Mandel stress~\cite{Epstein2000};
    $\lambda$ denote fiber stretch,
    $\vartheta$ denote growth multipliers;
    $\mathbf{f}_0$, $\mathbf{s}_0$, and $\mathbf{n}_0$ denote myocyte, sheet,
    and sheet-normal directions and $E_\bullet$ are strains in
    the respective direction;
    subscripts $\bullet_\mathrm{crit}$ denote physiologial limit levels,
    subscripts $\bullet_\mathrm{hom}$ denote homeostatic levels of the specific parameter;
    subscripts $\bullet_\mathrm{e}$ denote values with respect to the elastic
    deformation gradient $\mathbf{F}_{\rm{e}}$;
    $k(\bullet)$ are growth scaling functions.
}}%
  \vspace*{1ex}%
  \label{tab:KinGr}
  \resizebox{\linewidth}{!}{%
  \begin{tabularx}{1.12\linewidth}{>{\hsize=.3\hsize}L>{\hsize=.7\hsize}LL>{\hsize=.78\hsize}L}
  \toprule
  \textbf{Model} & \textbf{Geometry and Material} & \textbf{Driving Factor} & \textbf{Growth Laws} \\
  \midrule
  \citet{Kroon2009a} &
    Truncated ellipsoid; \newline transversely isotropic \cite{Kerc2003a} &
    Deviation of myofiber strain $s = \sqrt{E_{\mathrm{f}}+1} - 1$
    from its homeostatic level $s_\mathrm{hom}= 0.13$ &
    $\mathbf{F}_{\rm{g}}= {[\beta(s - s_\mathrm{hom})\Delta t +1]}^{1/3}\mathbf{I}$,
    \newline
    where $\beta$ is a rate constant \\
  \midrule
  \citet{Goek2010a}&
    Regularly shaped bi-ventricular model; isotropic material &
    Deviation from strain for eccentric growth \newline
    $\dot{\vartheta}^\parallel =
     k^\parallel({\vartheta}^\parallel)\left(\frac{1}{\vartheta^\parallel}\lambda
     - \lambda_\mathrm{crit}\right)$
    \newline
    Deviation from Mandel stress for concentric growth
    \newline
    $\dot{\vartheta}^{\perp} =
      k^\perp({\vartheta}^\perp)\left(\operatorname{tr}(\mathbf{M}_\mathrm{e})-
      p_{\rm{crit}}\right)$ &
    Eccentric growth: \newline
      $\mathbf{F}_{\rm{g}}= \mathbf{I}+[\vartheta^\parallel-1]\,\mathbf{f}_0\otimes\mathbf{f}_0$ \newline
      Concentric growth: \newline
      $\mathbf{F}_{\rm{g}}= \mathbf{I}+[\vartheta^{\perp}-1]\,\mathbf{s}_0\otimes\mathbf{s}_0$\\
  \midrule
  \citet{Rausch2011a} &
    Regularly shaped bi-ventricular model; orthotropic Holzapfel--Ogden~\cite{Holz2009c} &
    $\dot{\vartheta} = k(\vartheta)\left(\mathrm{tr}(\boldsymbol{\tau}) - p_{\rm{crit}}\right)$,
    with $\boldsymbol{\tau}$ the Kirchhoff stress tensor and
    $p_{\rm{crit}}$ the baseline pressure level &
    $\mathbf{F}_{\mathrm{g}}= \mathbf{I}+[\vartheta-1]\,\mathbf{s}_0\otimes\mathbf{s}_0$ \\
  \midrule
  \citet{Klepach2012a} &
    Patient-specific LV;\newline
    transversely isotropic Guccione~\cite{Gucc1991a} &
    Same as \citet{Rausch2011a}&
    $\mathbf{F}_{\rm{g}} = \vartheta \mathbf{f}_0 \otimes \mathbf{f}_0 + \frac{1}{\sqrt\vartheta}[\mathbf{I} - \mathbf{f}_0 \otimes \mathbf{f}_0]$ \\
  \midrule
  \citet{Kerckhoffs2012} &
    Thick-walled truncated ellipsoid;\newline
    transversely isotropic Guccione~\cite{Gucc1991a} &
    Stimulus for axial fiber growth $s_l = \max(E_{\rm{f}}) - E_{{\rm{f,set}}}$,
    and radial fiber growth $s_t = \min(E_\mathrm{cross,max}) -E_\mathrm{cross, set}$
    as differences between fiber strain $E_{\rm{f}}$ and maximum
       principal strain $E_\mathrm{cross,max}$ of the cross-sectional strain tensor:
       \(
         \mathbf{E}_{\rm{cross}} =
         \left[ E_{\rm{s}} \ E_{\rm{sn}};\ E_{\rm{sn}}\  E_{\rm{n}} \right]
       \)
       and set-points $\bullet_\mathrm{set}$
       &
    Transversely isotropic, incremental growth tensor,
       described by sigmoids and dependent on 10 parameters \\
  \midrule
  \citet{Lee2016} &
    Patient-specific LV;\newline
    transversely isotropic Guccione~\cite{Gucc1991a} &
    $\dot{\vartheta} =
      k(\vartheta,\lambda_{\rm{e}})(\lambda_{\rm{e}}-\lambda_\mathrm{hom})$ &
    $\mathbf{F}_{\rm{g}}= (\vartheta - 1) \mathbf{f}_0\otimes\mathbf{f}_0 + \mathbf{I}$
    \\
  \midrule
  \citet{Genet2016a} &
    Four-chamber human heart model; \newline
    orthotropic Guccione~\cite{Gucc1991a} &
    $\dot{\vartheta}=\frac{1}{\tau}\langle\lambda-\lambda_{\rm{crit}}\rangle$,
      where $\tau$ is a scaling parameter in time and $\langle\ \rangle$ Macaulay brackets&
    Eccentric growth:
    $\mathbf{F}_{\rm{g}} =
    \mathbf{I}+[\vartheta-1]\,\mathbf{f}_0\otimes\mathbf{f}_0$
    \newline
    Concentric growth:
      $\mathbf{F}_{\rm{g}}
        = \vartheta\mathbf{I}
        + [1 - \vartheta]\mathbf{f}_0\otimes\mathbf{f}_0$ \\
  \midrule
  \citet{Witzenburg2018} &
    LV treated as thin-walled spherical pressure vessel;
    time-varying elastance compartmental model &
    Same as \citet{Kerckhoffs2012} &
    Same as \citet{Kerckhoffs2012} \\
  \midrule
  \citet{DelBianco2018} &
    Truncated ellipsoid;\newline
    orthotropic Holzapfel--Ogden \cite{Holz2009c} &
    Local growth increments $\vartheta^n$ between cycles $n$ and $n+1$:
    $\vartheta^n = 1 + k (\prod\limits_{i=1}^{n-1}\vartheta^i)
    \left[\rm{tr}(\mathbf{M}_\mathrm{e})_\mathrm{n}
          - \rm{tr}(\mathbf{M}_\mathrm{e})_\mathrm{hom}\right]$ &
    Same as \citet{Goek2010a} for eccentric growth\\
  \midrule
  \citet{Peir2019a} &
  Subject specific LV;\newline
  orthotropic Holzapfel--Ogden \cite{Holz2009c} &
  $ \dot{\vartheta}^\parallel =
    \frac{1}{\tau}\langle\lambda_{\rm{e}}-\lambda_{\rm{crit}}\rangle$,
  $\dot{\vartheta}^{\perp} = 0$,
  where $\tau$ is a scaling parameter and $\langle\ \rangle$ Macaulay brackets&
  $\mathbf{F}_{\rm{g}} = \vartheta^\parallel \;[\mathbf{f}_0 \otimes \mathbf{f}_0] + \vartheta^{\perp}\;[\mathbf{I} - \mathbf{f}_0 \otimes \mathbf{f}_0]$ \\
  \bottomrule
  \end{tabularx}%
}
\end{table}
\end{landscape}

\begin{landscape}
\begin{table}[htpb]
  \caption{{\small Overview of studies since 2013 utilizing the
  constrained mixture theory (classical or hybrid) to study G\&R\@.}}%
  \vspace*{0.8em}%
  \label{tab:ConsGr}
  \resizebox{\linewidth}{!}{%
    \begin{tabularx}{1.45\linewidth}{>{\hsize=.5\hsize}LLLLLL>{\hsize=.48\hsize}L}
  \toprule
  \textbf{Study} & \textbf{Production Rate} $\rm{m}^i_{\tau}$ &
  \multicolumn{2}{c}{\textbf{Survival Function} $q^{i}(s,\tau)$} &
  \textbf{Remodeling} & \textbf{Volumetric Growth} & \textbf{Geometry}  \\
          &Collagen &  \centering Elastin &  \centering Collagen & & & \\
  \midrule
  \citet{Vale2013a} \newline[1em] Classical &
    Dependent on deviation from homeostatic stress and homeostatic shear stress. & &
    Exponential decay function &
    New deposition of collagen in direction of first and second principal
      directions of deviatoric part of Cauchy stress tensor. &
      Isotropic\newline[0.5em]
    $W_{\rm vol}(J(t)) = \mbox{$\qquad\frac{K(J(t))}{2}{(J(t) - J_\mathrm{his}(t))}^2$}$,\newline[0.5em]
      where $K(J(s))$ is a growth dependent penalty function approaching
      $\infty$ when $J(t) \to 0$ &
    2-layered aorta \\
  \midrule
  \citet{Eriksson2014} \newline[1em] Hybrid & &
      Axially constrained exponential degradation function &
      Dependent on deviation of collagen fiber stretch from
      homeostatic attachment stretch (given as material parameter) &
      Evolving recruitment stretches &
      Isotropic\newline[0.5em]
      \mbox{$W_{\rm vol} = \frac{\mu_{\kappa}}{2}(J(t)-J_{\rm his}(t))$} &
      2-layered aorta \\
  \midrule
  \citet{Wu2015} \newline[1em] Classical &
    Deviation of wall shear stress and collagen fiber stress from homeostatic
      value and basal value of mass production rate per collagen family & &
      Stepwise decrease normalized by the lifespan of collagen
      (\numrange{70}{80}~days)
      same as \cite{Figueroa2009} &
    Stretch ratio of newly produced collagen set to \num{1.05} current direction
      defined as $\mathbf{F}(t)\mathbf{e}^k_0$, where $\mathbf{e}^k_0$
      is the initial collagen fiber direction &
    membrane model, no volumetric growth &
    Patient-specific, infrarenal aorta \\
  \midrule
  \citet{Virag2015} \newline[1em] Classical &
    Driven by the deviation of the overall wall stress from homeostatic value &
    Exponential decay function depending on half-life of \num{40}~years and
      increased degradation due to inflammatory factors &
    Exponential decay function depending on ratio of current and homeostatic
      fiber tension and the presence of collagenases accelerating degradation &
    Constant pre-stretch for all constituents &
    No volumetric growth, semi-analytical solution &
    Axisymmetric, cylindrical geometry fusiform lesion \\
  \midrule
  \citet{Famaey2018} \newline[1em] Classical &
    Dependent on deviation of fiber stress from homeostatic
      value and basal production rate;
      Effects of shear wall stress neglected & &
    Exponential decay function dependent on the fiber tension
      of a certain cohort collagen fiber families and a homeostatic
      constant decay &
    Deposition stretch assumed to be known constant for all collagen fibers
      Elastin prestretch is iteratively found by pressurizing a load-free
      geometry and prescribing prestretches until the geometry matches the
      reference configuration &
    No volumetric growth &
    Single linear hexahedral element \\
  \midrule
  \citet{Grytsan2017} \newline[1em] Hybrid &
    Deviation of fiber stretch from homeostatic value &
    Exponential decay function for initiation of aneurysm
      development, as introduced by \cite{Watt2004a}  &
    Governed by rate constants and deviation of fiber stretch from target value &
    Evolving recruitment stretches &
    Three different growth tensors studies: isotropic, in-plane and in-thickness
      volumetric growth &
    Thick walled cylinder \\
  \midrule
  \citet{Lin2019} \newline[1em] Hybrid &
    Dependent on deviation of fiber stretch from homeostatic value &
    Non-axisymmetric degradation function, dependent on axial and
      circumferential location &
    Exponential degradation function dependent on time (half-life) and
      fiber stretch. &
    $\mathbf{F}$ of collagen fiber dependent on evolution of recruitment stretch &
    Isotropic growth
      $W_\mathrm{vol}(t) = \frac{1}{2}\kappa{(\frac{J(t)}{J(t^-)}-1)}^2$,
      where $J(t^-)$ is the volume change in previous time
      increment and $\kappa$ is a penalty parameter &
    Thick walled cylinder \\
  \midrule
  \citet{Horvat2019} \newline[1em] Classical &
    Deviation of wall shear stress and intramural Cauchy stress from
      homeostatic values &
    Exponential decay function for initiation of aneurysm development,
      as introduced by \cite{Watt2004a} for symmetric case, non-symmetric
      case additionally dependent on circumferential direction &
    Governed by rate constants and deviation of fiber stretch from target value. &
    Constant pre-stretch for all constituents &
    Isotropic growth
      $W_{\rm vol}(t) = \frac{1}{d_1}{(J - \frac{M(s)}{M(0)})}^2$,
      where $d_1$ is a penalty parameter, $M(0)$ and $M(s)$ are initial and
      targeted current total mass &
    3-layered, thick walled cylinder \\
  \bottomrule
  \end{tabularx}%
}
\end{table}
\end{landscape}
%
%%%%%%
%
%
%
%
\begin{landscape}
\begin{table}[tbp]
  \caption{\small{Summary table showing data collected from human biopsy
    samples in clinical studies researching aortic stenosis.
    Usage of different units than indicated in the top row is marked next to
    the value. AVR:~Aortic valve replacement, AS:~aortic sclerosis,
    $\diameter$:~diameter, EF:~ejection fraction, EDP:~LV end diastolic pressure,
    EDS:~LV end systolic pressure, HR:~heart rate, WT:~wall thickness.}}%
  \vspace*{1ex}%
  \label{tab:expTab}
  \resizebox{\linewidth}{!}{%
    \aboverulesep=0ex
    \belowrulesep=0ex
    \setlength{\extrarowheight}{1.0pt}
    \begin{tabularx}{1.38\linewidth}{ll|cccc|cccccc}
  \toprule
\textbf{Study}& & \textbf{Myocyte $\diameter$} & \multicolumn{3}{c|}{\textbf{Volume Fraction} [\si{\%}]} & \textbf{LV Mass} & \textbf{EF}
  & \textbf{EDP} & \textbf{ESP} & \textbf{HR} & \textbf{LV WT} \\
  & & [\si{\um}] & Myocytes & ECM & Collagen &  [\si{\g/\m\squared}] &
  [\si{\%}] & [\si{\mmHg}] & [\si{\mmHg}] & [\si{bpm}] & [\si{\cm}] \\
 \midrule
 \citet{Nitenberg1979} & AS &\num{31\pm9}& & &\num{39\pm6}& & \num{58} & & & & \\
 \midrule
 \citet{Hess1981a} &
 Control & \num{13.7\pm1}   & & &\num{2\pm1}& \num{79\pm5} &\num{69\pm4}&\num{12\pm1}&\num{116\pm3} &\num{78\pm3}& \\
    & AS & \num{26.8\pm1.7} & & &\num{15\pm1}&\num{171\pm16}&\num{62\pm5}&\num{18\pm3}&\num{196\pm11}&\num{74\pm6}& \\
 \midrule
 Schaper 1981 & &\num{23\pm8}& \num{44.9\pm14.3} & & \num{19.5\pm5.3} & & \num{65} & & & & \\
 \midrule
 \citet{Schwarz1981} &
 EF \SI{>55}{\%} & & \num{48.4\pm4.7} & & \num{16.3\pm5.5} & \num{148.3\pm20.9} &\num{65\pm10}& \num{21.4\pm7.2} & \num{195.4\pm10.9} & \num{70.9\pm8}&  \\
 & EF \SI{<55}{\%} & & \num{42.1\pm4.9} & & \num{14.7\pm3.8} & \num{199.8\pm43.5} & \num{44} & \num{28.7\pm6.6} & \num{205.5\pm44.4} & \num{82.6\pm12.4} &  \\
 \midrule
 Kunkel 1982 &
   EF \SI{> 50}{\%} & & \num{46.2\pm3.4} & & & & \num{\ge~65} & & & & \\
 & EF \SI{< 50}{\%} & & \num{27.9\pm9.3} & & & & \num{< 50}   & & & & \\
 \midrule
 \citet{Hess1984} &
       Control &\num{14\pm1}& & & \num{2\pm1}& \num{81\pm5}& \num{69\pm2}& \num{8\pm1}&\num{117\pm3} &\num{71\pm4}& \\
  & AS pre AVR &\num{31\pm1}& & &\num{15\pm1}&\num{188\pm16}&\num{58\pm5}&\num{19\pm3}&\num{210\pm11}&\num{75\pm3}& \\
 & AS post AVR &\num{26\pm1}& & &\num{26\pm3}&\num{118\pm11}&\num{62\pm4}&\num{15\pm2}&\num{150\pm5} &\num{71\pm3}& \\
 \midrule
 \citet{Huysman1989} &
 Control & & 73.4 & \num{26.6\pm8.0} & &\num{200\pm40}& & & & & \\
 & AS subendocardial & & 61.8 & \num{38.2\pm8.7} & &\num{324\pm71}& & & & \multicolumn{1}{l}{} & \multicolumn{1}{l}{} \\
   & AS subepicardial & & 59.8 & \num{40.2\pm6.8} & & & & & & \multicolumn{1}{l}{} & \multicolumn{1}{l}{} \\
 \midrule
 \citet{Krayenbuehl1989} &
    Control & \num{21.2\pm2.0} & \num{57.2\pm2.6} & & \num{7.0\pm1.8} & & & & & \multicolumn{1}{l}{} & \multicolumn{1}{l}{} \\
  & Pre AVR & \num{30.9\pm4.7} & \num{57.7\pm5.9} & & \num{18.2\pm6.2} &\num{186\pm52}&\num{59\pm15}& \num{18.5\pm8.7} &\num{206\pm32}& \multicolumn{1}{l}{} & \num{1.25\pm0.18} \\
  & intermediate post AVR &\num{28\pm3.6} & \num{56.8\pm4.8} & & \num{25.8\pm8.7} &\num{115\pm28}&\num{65\pm10}& \num{12.5\pm5.1} &\num{144\pm19}& & \num{1.00\pm0.18} \\
   & 18 Months post AVR & \num{28.7\pm4.4} & \num{49.0\pm5.9} & & \num{13.7\pm3.6} &\num{94\pm20}&\num{57\pm16}& \num{12.1\pm3.2} &\num{138\pm14}& & \num{0.87\pm0.12} \\
 \midrule
 \citet{Vliegen1991} &
   Control    & & \num{81.22} & \num{18.78} & &\num{189\pm27}& & & & & \\
   & Moderate & & \num{79.02} & \num{20.98} & &\num{274\pm22}& & & & & \\
   & Severe   & & \num{78.24} & \num{21.66} & &\num{408\pm24}& & & & & \\
 \midrule
 \citet{Villari1993} &
   Control   & \num{21.2\pm2,0} & \num{57.2\pm2.6} & & \num{1.6\pm0.9} &\num{85\pm14} &\num{66\pm3} &\num{11\pm2} &\num{130\pm14}&\num{75\pm12}& \\
   & Group 1 & \num{28.1\pm3.9} & \num{55.9\pm4.3} & & \num{7.6\pm2.8} &\num{171\pm34}&\num{69\pm8} &\num{26\pm9} &\num{191\pm43}&\num{76\pm10}& \\
   & Group 2 & \num{28.3\pm4.8} & \num{55.1\pm2.7} & & \num{2.2\pm0.2} &\num{169\pm27}&\num{52\pm13}&\num{25\pm12}&\num{181\pm30}&\num{82\pm14}& \\
   & Group 3 & \num{28.4\pm2.9} & \num{55.5\pm3.7} & & \num{9.6\pm4.2} &\num{187\pm30}&\num{53\pm14}&\num{31\pm11}&\num{187\pm36}&\num{74\pm13}& \\
 \midrule
 \citet{Villari1995} &
 Control     & \num{21.2\pm2.0}&             &  &\num{7\pm2} &\num{86\pm10} &\num{64\pm4} &\num{12\pm2}&\num{118\pm13}&\num{72\pm11}&\num{0.78\pm0.05} \\
   & Pre AVR & \num{33\pm4}   &             &  &\num{16\pm5}&\num{202\pm41}&\num{55\pm11}&\num{20\pm7}&\num{202\pm23}&\num{76\pm12}&\num{1.27\pm0.11} \\
   & $22\pm 8$ Months Post AVR &\num{29\pm4}&& &\num{28\pm8}&\num{137\pm32}&\num{60\pm9} &\num{14\pm5}&\num{138\pm14}&\num{75\pm6} &\num{1.05\pm0.13} \\
   & $81\pm 11$ Months Post AVR&\num{28\pm3.6}&&&\num{13\pm2}&\num{116\pm23}&\num{60\pm10}&\num{14\pm4}&\num{135\pm20}&\num{76\pm7} &\num{0.96\pm0.15} \\
 \midrule
 \citet{Akdemir2002} &
 Control & & & & \num{1.86}\;(\numrange{1.62}{2.64}) &\num{119\pm49}&\num{66\pm8.6} &\num{7\pm2.8} & & & \\
    & AS & & & & \num{2.46}\;(\numrange{1.52}{3.92}) &\num{196\pm41}&\num{61\pm12.9}&\num{15\pm9.0}& & & \\
 \midrule
 \citet{Hein2003a} &
 Control           &\num{424\pm37}\,\si{\um\squared} &\num{380\pm20}\,\si{\npm} & & \num{12.00} &\num{104\pm14}&\num{61\pm8}&\num{8\pm1}&\num{130\pm17}& & \\
 & EF \SI{>50}{\%} &\num{561\pm97}\,\phantom{\si{\um\squared}}&\num{366\pm45}\,\phantom{\si{\npm}}& & \num{30.00} &\num{137\pm26}&\num{59\pm8}&\num{15\pm5}&\num{191\pm25}& & \\
 & EF \SIrange{30}{50}{\%} &\num{658\pm84}\,\phantom{\si{\um\squared}}&\num{290\pm21}\,\phantom{\si{\npm}}& & \num{30.00} &\num{131\pm57}&\num{41\pm5}&\num{18\pm6}&\num{180\pm15}& & \\
 & EF \SI{<30}{\%} &\num{593\pm64}\,\phantom{\si{\um\squared}}&\num{216\pm32}\,\phantom{\si{\npm}}& & \num{40.00} &\num{153\pm35}&\num{24\pm5}&\num{24\pm5}&\num{156\pm15}& & \\
 \midrule
 \citet{Herrmann2011} &
 Moderate AS & & & & &\num{132\pm37}&\num{64\pm8}& &\num{141\pm19}& & \num{0.43\pm0.04} \\
   & Severe AS, High Gradient & \num{12.2\pm1.3} & & & \num{1.8\pm0.8} &\num{195\pm53}&\num{55\pm9}& &\num{123\pm16}& & \num{0.56\pm0.1} \\
   & Severe AS, EF \SI{>50}{\%} & \num{13.1\pm1.5} & & & \num{3.9\pm0.6} &\num{162\pm39}&\num{61\pm5}& &\num{130\pm20}& & \num{0.58\pm0.1} \\
   & Severe AS, EF \SI{<50}{\%} & \num{13.7\pm1.3} & & & \num{4.8\pm0.6} &\num{197\pm35}&\num{36\pm10}& &\num{128\pm24}& & \num{0.44\pm0.08} \\
 \midrule
 \citet{Treibel2018} &
 Pre AVR           & & \num{71.8} & \num{28.2\pm2.9} & \num{7.7}\;(\numrange{4.2}{12.7}) &\num{88\pm26}&\num{71\pm16}& & & & \\
 & 1 year post AVR & & \num{70.1} & \num{29.9\pm4}   &                       &\num{71\pm19}&\num{74\pm12}& & & & \\
 \bottomrule
 \end{tabularx}
 }
\end{table}
\end{landscape}

\begin{landscape}
\begin{table}[tbp]
  \caption{Summary of the main changes occurring in myocytes, collagen, fibroblasts and inflammatory cells
  following pressure overload. EF: Ejection fraction, ECM: Extracellular matrix, MMPs: Matrix metalloproteinases}%
  \vspace*{1ex}%
  \label{tab:pathogenesisTable}
  \resizebox{\linewidth}{!}{%
    \aboverulesep=0ex
    \belowrulesep=0ex
    \setlength{\extrarowheight}{3.0pt}
    %\begin{tabularx}{1.\linewidth}{>{\hsize=1.3\hsize}L>{\hsize=1.7\hsize}L>{\hsize=1.78\hsize}L>{\hsize=1.78\hsize}LL}
    \begin{tabularx}{1.2\linewidth}{>{\hsize=0.6\hsize}LLLL>{\hsize=0.8\hsize}L>{\hsize=0\hsize}L}
  \toprule
  \textbf{Time Point} & \textbf{Myocytes} & \textbf{Collagen} & \textbf{Fibroblasts} & \textbf{Inflammatory Cells} \\
  \midrule
  Immediate reaction\linebreak $<$ 2 weeks
  &
  Synthesis rises; area of transverse tubuli rises \textcolor{review2}{\cite{Weber1987, Ferr2002a}};\linebreak
  $\Rightarrow$ hypertrophy,
  increase in diameter, not in length \textcolor{review2}{\cite{Kajstura2004}}.
  &
  Type III synthesis rises (thin fibers) \textcolor{review2}{\cite{Webe1989a}};\linebreak
  degradation rises.
  & & \\
  \\[-1em]
  & \multicolumn{4}{l}{\cellcolor{gray!20}\bf Decline in relative collagen content due to myocyte hypertrophy \textcolor{review2}{\cite{Bonnin1981}}}
  \\ \\[-1em]
  \midrule
  Mid term \linebreak
  \numrange{2}{4} weeks
  &
  Synthesis stabilized \textcolor{review2}{\cite{Webe1989a}}.
  &
  Synthesis rises \numrange{6}{8} times up to \SI{4}{\%} per day;
  synthesis exceeds degratation;
  degradation returns to baselevel. \textcolor{review2}{\cite{Webe1989a}}
  &
  Collagen synthesis carried out by existing fibroblasts \textcolor{review2}{\cite{Akdemir2002}}.
  & \\
  \\[-1em]
  & \multicolumn{4}{l}{\cellcolor{gray!20}\bf Fibrosis without cell necrosis:
  reactive fibrosis. \textcolor{review2}{\cite{Weber2013}} Hypothesis: reverse remodelling still possible. }\\
  \\[-1em]
  \midrule
  Long term \linebreak
  $>$ 4 weeks
  &
  Get ensnared in dense collagen meshwork \textcolor{review2}{\cite{Webe1989a}}.
  &
  Synthesis rises 3 times up to \SI{2}{\%} per day \textcolor{review2}{\cite{Akdemir2002}};
  expanded intermuscular spaces filled with perimysial collagen \textcolor{review2}{\cite{Akdemir2002}};
  meshwork around myocytes increases \textcolor{review2}{\cite{Gabb2003a}};
  rise around intra-myocardial coronary arteries \textcolor{review2}{\cite{Gabb2003a}}.
  &
  Proliferation occurs, trigger collagen formation \textcolor{review2}{\cite{VanPutten2016}}.
  & \\
  \\[-1em]
  & \multicolumn{4}{l}{\cellcolor{gray!20}\bf Cell necrosis: adaptive fibrosis \textcolor{review2}{\cite{Weber2013}}.
  Hypothesis: reverse remodelling no longer possible} \\
  \\[-1em]
  & Cell death occurs \textcolor{review2}{\cite{Hein2003a}}.
  &
  Lost myocytes replaced by type I collagen \textcolor{review2}{\cite{Gabb2003a}}.
  &
  Proliferation into myofibroblasts \textcolor{review2}{\cite{VanPutten2016}};
  expression of $\alpha$-actin, closing spaces vacant after myocyte death \textcolor{review2}{\cite{Gabb2003a}};
  scar formation \textcolor{review2}{\cite{Gabb2003a}}.
  &
  Migration to cell death sites;
  expression of MMPs which degenerate ECM. \textcolor{review2}{\cite{Leonard2012}}
  \\
  \\[-1em]
  & \multicolumn{4}{l}{\cellcolor{gray!20}\bf Diastolic Heart Failure:
  EF preserved, capability to increase cardiac output when needed impaired. \textcolor{review2}{\cite{LeGrice2012b}}} \\
  \\[-1em]
  &
  Further cell death due to decrease in capillary density
  and inhibited capability to generate force during systole. \textcolor{review2}{\cite{Hein2003a,Weber2013}}
  &
  Ensnare myocytes, inhibit stretching during diastole;
  degradation rises due to MMPs from inflammatory sites. \textcolor{review2}{\cite{Gabb2003a}}
  &
  Invade from adventitia of vessels. \textcolor{review2}{\cite{VanPutten2016}}
  &
  Invade from adventitia of vessels;
  secretion of MMPs, degrading collagen. \textcolor{review2}{\cite{Leonard2012}}
  \\
  \\[-1em]
   & \multicolumn{4}{l}{\cellcolor{gray!20}\bf Ongoing dilatation of the
   myocardium due to cell death and collagen degradation by MMPs,} \\
   & \multicolumn{4}{l}{\cellcolor{gray!20}\bf resulting eventually in
   Systolic Heart Failure \textcolor{review2}{\cite{Weber2013}}} \\
  \\[-1em]
  \bottomrule
 \end{tabularx}
 }
\end{table}
\end{landscape}
\end{document}